\documentclass[a4paper,final]{article}

\usepackage[numbers,square,sort&compress]{natbib}
\usepackage{color}
\usepackage{todonotes}
\usepackage{ifthen,xspace}
\usepackage{amsmath}
\usepackage{amssymb}
\usepackage{multirow}
\usepackage{stmaryrd}
\usepackage{xcolor}
\usepackage[colorlinks=true,citecolor = green!60!black,urlcolor = blue!60!black,linkcolor = red!60!black]{hyperref}

\usepackage[top=2.cm, bottom=2.5cm, left=2.5cm, right=2.5cm]{geometry}

\DeclareMathOperator*{\argmin}{arg\,min}
\DeclareMathOperator*{\argmax}{arg\,max}
\graphicspath{{graphics/}}

\title{Quantifying Emergent Behavior of Autonomous Robots}

\author{Georg Martius$^{1}$* and Eckehard Olbrich$^{2}$\\[.5em]
\normalsize $^{1}$\,IST Austria, Am Campus 1, 3400 Klosterneuburg, Austria\\
\normalsize $^{2}$\,Max Planck Institute for Mathematics in the Sciences, Inselstr. 22, 04103 Leipzig, Germany
}

\begin{document}

\maketitle{}

\begin{abstract}
Quantifying behaviors of robots which were generated autonomously from task-independent objective functions is an important prerequisite for objective comparisons of algorithms
 and movements of animals.
The temporal sequence of such a behavior can be considered as a time series and hence complexity measures developed for time series are natural candidates for its quantification. The predictive information and the excess entropy are such complexity measures. They measure the amount of information the past contains about the future and thus quantify the nonrandom structure in the temporal sequence.
However, when using these measures for systems with continuous states one has to deal with the fact that their values will depend on the resolution with which the systems states are observed. For deterministic systems both measures will diverge with increasing resolution.
We therefore propose a new decomposition of the excess entropy in resolution dependent and resolution independent parts and discuss how they depend on the dimensionality of the dynamics, correlations and the noise level. For the practical estimation we propose to use estimates based on the correlation integral instead of the direct estimation of the mutual information using the algorithm by Kraskov et al.~(2004) which is based on next neighbor statistics because the latter allows less control of the scale dependencies.
Using our algorithm we are able to show how autonomous learning generates behavior of increasing complexity with increasing learning duration.
\end{abstract}

\newcommand{\ie}{i.\,e.\xspace}
\newcommand{\eg}{e.\,g.\xspace}
\newcommand{\wrt}{w.\,r.\,t.\xspace}
\newcommand{\Eqn}[1]{Equation~\eqref{#1}}
\newcommand{\eqn}[1]{Eq~\eqref{#1}} 
\newcommand{\eqns}[2]{Eqs~(\ref{#1}, \ref{#2})} 
\newcommand{\eqnr}[2]{Eqs~(\ref{#1}--\ref{#2})} 
\newcommand{\eqnp}[1]{(\ref{#1})} 
\newcommand{\Fig}[1]{Fig~\ref{#1}}
\newcommand{\fig}[1]{Fig~\ref{#1}}
\newcommand{\infig}[1]{{\bf #1}}
\newcommand{\tab}[1]{Table~\ref{#1}}
\renewcommand{\sec}[1]{Sec.~\ref{#1}}

\newcommand{\PI}{\ensuremath{\operatorname{PI}}}
\newcommand{\KSG}{\textrm{KSG}}
\newcommand{\IKSG}[1]{I_\KSG^{(#1)}}
\newcommand{\Renyi}{R\'enyi}
\newcommand{\htwo}{h^{(2)}}
\newcommand{\Htwo}{H^{(2)}}
\newcommand{\Dtwo}{D^{(2)}}
\newcommand{\Etwo}{E^{(2)}}

\newcommand{\EEState}{\ensuremath{E_{\textrm{state}}}}
\newcommand{\EEMiddle}{\ensuremath{E_{\eps}}}
\newcommand{\EEMem}{\ensuremath{E_{\textrm{mem}}}}
\newcommand{\EECore}{\ensuremath{E_{\textrm{core}}}}

\newcommand{\eps}{\ensuremath{\varepsilon}}
\newcommand{\const}[0]{\ensuremath{\textrm{const}}}

\newcommand{\Real}{\ensuremath{\mathbb R}}        
\newcommand{\T}{\ensuremath{\top}}                

\newcommand{\Hexapod}[0]{\textsc{Hexapod}\xspace}
\newcommand{\Snake}[0]{\textsc{Snake}\xspace}

\newcommand{\Geo}[2][]{\todo[#1,color=yellow!60,size=\scriptsize]{#2}\xspace}
\newcommand{\Eck}[2][]{\todo[#1,color=green!60,size=\scriptsize]{#2}\xspace}

\newlength{\pich}
\newcounter{subplot}
\renewcommand{\thesubplot}{(\alph{subplot})}
\newcommand{\subplot}[1]{\refstepcounter{subplot}\thesubplot \label{#1}}
\newcommand{\resetsubplot}{\setcounter{subplot}{0}}

\section{Introduction}

Whenever we create behavior in autonomous robots we strive for a suitable measure in order
 to quantify success, learning progress or compare  algorithms with each other.
When a specific task is given, then typically the task provides a natural measure of success,
for instance walking may be measured by velocity and perturbation stability.
In cases where behavior is learned via optimization of a global objective function then the
same function can also be used as a quantification,
so creating and quantifying behavior often go hand in hand. 
This also applies in principle to behavior from task-independent\footnote{A task is here understood as a specific desired target behavior such as walking or reaching.} objectives that have been
 recently more and more successful in generating emergent autonomous behavior in robots~\cite{ay08:predinf_explore_behavior,ZahediAyDer2010:HigherCoordination,Klyubin2005:Empowerment,SalgeGlackinPolani2014:EmpowermentIntro,schmidhuber:02predictable,oudeyer:05c,FrankSchmidhuber2013:curiosity}.
However, there are several cases of emergent behavior where this strategy fails:
if the behavior arises from
a) optimizing a local function~\cite{DerMartius11,MartiusDerAy2013}, b) optimizing a crude approximation of a computationally expensive objective function~\cite{MartiusDerAy2013}, c) local interaction rules without an explicit optimization function~\cite{Sayama2014:GSO-swarms,DerMartius13,DerMartius2015:DEP},
and d) a biological system (\eg freely moving animals) where we don't know the underlying optimization function.
Thus, independent of the origin of behavior it would be useful to have a quantitative description of its structure.
This would allow to objectively compare algorithms with each other and to compare technical with biological  autonomous systems.
This paper presents a measure of behavioral complexity that is suitable for the analysis of this kind of emergent behavior.

We base our measure on the predictive information~(PI)~\citep{bialek01} of a set of observables, such as joint angles.
The PI is the mutual information between past and future of a time-series and captures
 how much information (in bits) can be used from the past to predict the future.
It is also closely linked to the excess entropy~\citep{shaw84,Crutchfield03} or effective measure complexity~\citep{Grassberger86}, and it is a natural measure for the complexity of a time series because it provides a lower bound for the information necessary for optimal prediction.
Thus it is a promising choice as a measure.
Given these favorable properties, the PI was also proposed as an intrinsic drive for behavior generation
 in autonomous robots~\citep{ay08:predinf_explore_behavior,ZahediAyDer2010:HigherCoordination,MartiusDerAy2013}.
On an intuitive level maximizing the PI of the sensor process,
leads to a high variety of sensory input --- large marginal entropy --- while keeping a temporal structure --- high predictability corresponding to a small entropy rate.

Unfortunately, there are conceptual and practical challenges in using the PI for generating and measuring autonomous behavior.
Conceptually, for systems with continuous values the PI provides a family of functions depending on the resolution of the measurements and
practically it is difficult to estimate the PI.
For a fixed partition (single resolution) in a low-dimensional case it is possible
to estimate the one-step PI and adapt a controller to maximize it~\citep{ZahediAyDer2010:HigherCoordination}.
In high-dimensional systems it cannot be used directly in an online algorithm.
An alternative approach~\citep{MartiusDerAy2013} uses a dynamical systems model,
 linearization and a time-local version of the PI~\cite{MartiusDerAy2013} to obtain an explicit gradient for locally maximizing PI for continuous and high-dimensional state spaces (no global optimization).
A simplified version of the resulting neural controller~\cite{DerMartius13}  was used for generating the examples presented in this paper.

The aim of the present paper is to provide a measure of behavioral complexity that can be applied
to emergent autonomous behavior, animal behavior and potentially other time-series.
In order to do this it turns out that the predictive information as a complexity measure has to be
refined by taking into account its scale dependency.
Because the systems of interest are high-dimensional
we investigate different methods to estimate information theoretic quantities
in a resolution dependent way and present results for different robotic experiments.
The emergent behavior in these experiments are generated by the above mentioned learning rules.
On the one hand we want to increase our understanding of the learning process guided by the time-local version of the PI and on the other hand demonstrate how
the resolution dependence allows us to quantify behavior on different length-scales.

We argue that many natural behaviors are low-dimensional, at least on a coarse scale.
For instance in walking all joint angles can be typically related to a single phase, so in the extreme case it is one-dimensional~\citep{vanderWeele01:modeiteraction}.
However, as we show below in \ref{ss:EntrDimExcessEnt}, this contradicts to the global maximization of the PI which would maximize the dimension.
We discuss this contradiction and how it was circumvented in the practical applications.
Along with the paper there is a supplementary material located at \href{http://playfulmachines.com/QuantBeh2015/}{\tt{\small playfulmachines.com/QuantBeh2015}} and the source code can be found in the repository \href{https://github.com/georgmartius/behavior-quant}{\tt{\small github.com/georgmartius/behavior-quant}}.

The plan of the paper is as follows. In \sec{ss:EntrDimExcessEnt} we introduce the information theoretic quantities and in \sec{subsec:methods-estimation} the two estimation methods used in the paper. In \sec{sec:decomp:algo} we describe the algorithm used to calculate the quantification measures. In \sec{ss:Example} we demonstrate their behavior at the example of the Lorenz system including the effects of noise.
\sec{s:Results}  starts with explaining the robots and the control framework used for the experiments which are described in \ref{Experiments}. The results of the experiments quantification are presented in \sec{s:quant-behavior} and are discussed in \sec{s:Discussion}.

\section{Methods}\label{sec:Methods}

\subsection{Entropy, dimension and excess entropy}\label{ss:EntrDimExcessEnt}

Starting point is a (in general) vector valued stationary time series $\{y_0,y_1,y_2,\dots,y_N\}$. It is used to reconstruct the phase space of the underlying dynamical system by using delay embedding: $\vec x _t=\{y_t,y_{t-\tau},y_{t-2\tau},\dots,y_{t-(m-1)\tau}\}$.
If the original time series was $d-$dimensional the reconstructed phase space will be of dimension $d \times m$ \footnote{There are more general embeddings possible, e.g. involving different delays or different embedding dimensions for each component of the time series.}. In the following, we will only consider the case $d=1$, i.e. scalar time series. The generalization to vector-values time series is straight forward.
Let us assume that we are able to reconstruct approximately the joint probability distribution $p(\vec x _t, \vec x _{t-1}, ...)$ in a reconstructed phase space. Then we can characterize its structure using measures from information theory. Information theoretic measures represent a general and powerful tool for the characterization of the structure of joint probability distributions \citep{Grassberger91_2012,prokopenko09:ITprimer}. The uncertainty about the outcome of a single measurement of the state, i.e. about $\vec x _t$ is given by its \emph{entropy}. For discrete-valued random variable $X$ with values $x \in \cal{X}$ and a probability distribution $p(x)$ it is defined as
\begin{equation*}
H(X)=-\sum_{x \in \cal{X}} p(x) \log p(x) \;.
\end{equation*}
An alternative interpretation for the entropy is the average number
 of bits required to encode a new measurement.
In our case, however, the $y_t$ are considered as continuous-valued observables (that are measured with a finite resolution). For continuous random variables with a probability density $\rho(x)$ one can also define an entropy, the \emph{differential entropy}
\begin{equation}
H^C(X)=-\int \rho(x) \log \rho(x) dx\label{eqn:entropy:diff} \;.
\end{equation}

However, it behaves differently than its discrete counterpart: It can become negative and it will get even minus infinity if the probability measure for $X$ is not absolutely continuous \wrt to the Lebesgue measure -- for instance, in the case of the invariant measure of a deterministic system with an attractor dimension smaller than the phase space dimension.
Therefore, when using information theoretic quantities for characterizing dynamical systems researchers often prefer using the entropy for discrete-valued random variables.
In order to use them for dynamical systems with continuous variables
 usually either partitions of the phase space or entropy-like quantities based on coverings are employed.
These methods do not explicitly reconstruct the underlying invariant measure,
 but exploit the neighbor statistics directly.
Alternatively one could use direct reconstructions using kernel density
 estimators~\cite{Rosenblatt1956:KDE} or methods based on maximizing relative entropy~\cite{Shore1980:MaxEntr,Giffin2009:MrE} to gain parametric estimates.
These reconstructions, however, will always lead to probability densities, and are not suitable
 for reconstructing fractal measures which appear as invariant measures of deterministic systems.

In this paper we use estimators based on coverings, \ie correlation entropies \citep{Takens1998} and nearest neighbors based methods~\cite{KraskovSG04:EstimatingMI} considered in \ref{subsec:methods-estimation} below.
For the moment let us consider a partition of the phase space into hypercubes with side-length $\epsilon$.  For a more general definition of an $\epsilon$-partition \cite[see][]{Gaspard93c}. In principle one might consider scaling the different dimensions of $y_t$ differently, but for the moment we assume that the time series was measured using an appropriate rescaling. The entropy of the state vector $\vec x _t$ observed with a $\eps$-partition will be denoted in the following as $H(\vec X;\eps)$ with $\epsilon$ parameterizing the resolution.

How does the entropy change if we change $\eps$? The uncertainty about the potential outcome of a measurement will increase if the resolution of the measurement is increased, because of the larger number of potential outcomes. If $\vec{x}$ is a m-dimensional random variable and distributed according to a corresponding probability density function $\rho(x)$ we have asymptotically for $\epsilon \to 0$ (\cite[see Cover and Thomas][ch.8, p.248, theorem 8.3.1]{Cover2006} or \cite{Gaspard93c}).
\begin{equation}
H(X;\eps) \approx H^C(X)-m \log \eps \;.
\label{eps-entr-m}
\end{equation}
This is what we would expect for a stochastic system. However, if we observe a deterministic system the behavior of an observable depends how its dimension relates to the attractor dimension. If the embedding dimension is smaller than the attractor dimension the deterministic character will not be resolved and Eq.~\ref{eps-entr-m} still applies. However, if the embedding dimension is sufficiently high ($m > D$~\cite{Ding93}) then instead of a density function $\rho(x)$ we have to deal with a $D$-dimensional measure $\mu(x)$ and the entropy will behave as
\begin{equation}
H(X;\eps) \approx \const - D \log \eps \;.
\label{eps-entr-D}
\end{equation}
If an behavior such as in Eqs.~(\ref{eps-entr-m}) or (\ref{eps-entr-D}) is observed for a range of $\epsilon$ values we will call this range a \emph{stochastic} or \emph{deterministic scaling range}, respectively.

\paragraph{Conditional entropy and mutual information}
 Let us consider two discrete-valued random variables $X$ and $Y$ with values $x \in \cal{X}$ and $y \in\cal{Y}$, respectively. Then the uncertainty of a measurement of $X$ is quantified by $H(X)$. Now we might ask, what is the average remaining uncertainty about $X$ if we have seen already $Y$? This is quantified by the \emph{conditional entropy}
\begin{equation}
H(X|Y)=H(X,Y)-H(Y) \;.
\end{equation}
The reduction of uncertainty about $X$ knowing $Y$ is the information that $Y$ provides about $X$ and is called the \emph{mutual information} between $X$ and $Y$
\begin{equation}
I(X:Y)=H(X)-H(X|Y) \;.
\end{equation}

Having defined the $\eps$-dependent state entropy $H(\vec X;\eps)$ we can now ask, how much information the present state contains about the state of the system at the next time step. The answer is given by the mutual information between $\vec X _t$ and $\vec X _{t+1}$:
\begin{equation}
I(\vec X _t : \vec X _{t+1};\eps)=H(\vec X _{t+1};\eps)-H(\vec X _{t+1}|\vec X _t;\eps) \;.
\end{equation}

Using Eq.~\ref{eps-entr-m} one see, that for stochastic systems the mutual information will remain finite In the limit $\eps \to 0$ and can be expressed by the differential entropies:
\begin{equation}
I(\vec X _t : \vec X _{t+1}):=\lim_{\eps \to 0} I(\vec X _t : \vec X _{t+1};\eps) = H^C(\vec X _{t+1})-H^C(\vec X _{t+1}|\vec X _{t})
\end{equation}
Note, that this mutual information is invariant with respect to coordinate transformation of the system state, i.e.  if $\vec z_j=f(x_j)$  is a continuous and invertible function, then
\begin{equation}
I(\vec X _t : \vec X _{t+1})=I(\vec Z _t : \vec Z _{t+1})  \;.
\end{equation}
However, in the case of a deterministic system, the mutual information will diverge
\begin{equation}
I(\vec X _t : \vec X _{t+1};\eps) \propto -D \log \eps \qquad \mbox{for} \; \eps \to 0 \;.
\end{equation}
This is reasonable behavior because in principle the actual state contains an arbitrary large amount of information about the future. In practice, however, the state is known only with a finite resolution determined by the measurement device or the noise level.

\paragraph{Predictive information, excess entropy and entropy rate}

The unpredictability of a time series can be characterized by the conditional entropy of the next state given the previous states. In the following we will use an abbreviated notation for these conditional entropies and the involved entropies:
\begin{align}
H_m(\vec X_t;\eps) &:=H(\vec X_t,\vec X_{t-1}, \ldots, \vec X_{t-m+1};\eps) \label{eqn:Hm}   \\
h_m(\vec X_t;\eps) &:=H(\vec X_t|\vec X_{t-1}, \ldots, \vec X_{t-m};\eps) \label{eqn:hn}  \\
h_0(\vec X_t;\eps)&:=H(\vec X_t;\eps) \nonumber \;.
\end{align}
The \emph{entropy rate} (\cite[see Cover and Thomas][ch 4.2]{Cover2006}) is this conditional information if we condition on the infinite past
\begin{equation}
h_\infty(\vec X_t;\eps)=\lim_{m \to \infty} h_m(\vec X_t;\eps) \:
\end{equation}

In the following we assume stationarity, i.e. we have no explicit time dependence of the joint probabilities and therefore also of the entropies. Moreover, if it is clear from the context, which stochastic process is considered, we will write $H_m(\eps)$ and $h_m(\eps)$ instead of $H_m(\vec X_t;\eps)$ and $h_m(\vec X_t;\eps)$, respectively and it holds
\begin{equation}
h_m(\eps)=H_{m+1}(\eps)-H_m(\eps)  \;.
\label{eq:cond_entr}
\end{equation}

For deterministic systems the entropy rate will converge in the limit $\eps \to 0$ to the Kolmogorov-Sinai (KS-)entropy \cite{Cohen1985,Sinai2009} which is a dynamical invariant of the system  in the sense that it is independent on the specific state space reconstruction. Moreover, already for finite $m \ge D$,  $h_m(\eps)$ will not depend on $\eps$ for sufficiently small $\eps$ because of (\ref{eps-entr-D}).

To quantify the amount of predictability in a state sequence one might consider subtracting the unpredictable part from the total entropy of a state sequence. By doing this one ends up with a well known complexity measure for time series, the \emph{excess entropy} \citep{shaw84,Crutchfield03} or \emph{effective measure complexity} \citep{Grassberger86}
\begin{equation}
E(\eps)=\lim_{m \to \infty} E_m(\eps)
\label{limEm}
\end{equation}
with
\begin{equation}
E_m(\eps):=H_m(\eps)-m h_{m-1}(\eps) \;. \label{eqn:Em}
\end{equation}
The excess entropy provides a lower bound for the amount of information necessary for an optimal prediction. For deterministic systems, however, it will diverge because $H_m(\epsilon)$ will behave according to $(\ref{eps-entr-D})$ and $h_{m-1}(\eps)$ will become $\epsilon$-independent for sufficiently large $m$ and small $\epsilon$, respectively
\begin{equation}
E_m(\eps) \approx \const -D \log \eps \;,
\label{eqn:Em-det}
\end{equation}
with $D$ being the attractor dimension.

The \emph{predictive information}~\citep{bialek01} is the mutual information between the semi-infinite past and the future time series
\begin{equation}
\PI(\eps)=\lim_{m \to \infty} \PI_m(\eps)
\label{limPIn}
\end{equation}
with
\begin{align}
\PI_m(\eps)&:=I(\vec X_{t+m},\ldots,\vec X_{t+1};\vec X_t,\ldots,\vec X_{t-m+1};\eps) \label{eqn:PI} \\
&= 2 H_m(\eps)-H_{2m}(\eps) \nonumber
\end{align}
If the limits~(\ref{limEm}) and~(\ref{limPIn}) exist the predictive information $\PI(\eps)$ is equal to the excess entropy $E(\eps)$. For the finite time variants in general $\PI_m(\eps) \neq E_m(\eps)$:
\begin{equation}
\PI_m=E_m+\sum_{k=m}^{2m-1} (h_{m-1}-h_k) \ge E_m \;. \nonumber
\end{equation}
However, if the system is Markov of order $p$ the conditional probabilities will only depend on the previous $p$ time steps, $p(\vec{X_t}|X_{t-1},\ldots,X_{t-\infty})=p(\vec{X}_t|X_{t-1},\ldots,X_{t-p})$, hence $h_m=h_{m-1}$ for $m>p$ and therefore $E=E_{p+1}=\PI_{p+1}$.

\subsection{Decomposing the excess entropy for continuous states}
\label{sec:decomp}

In the literature \cite{Grassberger86,Crutchfield03,bialek01} both the excess entropy and the predictive information were studied only for a given partition --- usually a generating partition. Thus, using the excess entropy as a complexity measure for continuous valued time series has to deal with the fact that its value will be different for different partitions --- even for different generating ones.

In (\ref{eqn:Em-det}) we have seen that the excess entropy for deterministic systems becomes infinite in the limit $\eps \to 0$. The same applies to the predictive information (\ref{eqn:PI}).
Moreover, we have seen that the increase of these complexity measures with decreasing $\eps$ is controlled by the attractor dimension of the system. Does this means that in the case of deterministic systems the excess entropy as a complexity measure reduces to the attractor dimension? Not completely. The constant in (\ref{eqn:Em-det}) reflects not only the scale of the signal, but also statistical dependencies or memory effects in the signal, in the sense that it will be larger if the conditional entropies converge slower towards the KS-entropy.

How can we separate the different contributions? We will start by rewriting \eqn{eqn:Em} as a sum. Using the conditional entropies \eqnp{eq:cond_entr} we get
\begin{equation*}
E_m(\eps)=\sum_{k=0}^{m-1} (h_k(\eps)-h_{m-1}(\eps)) \;.
\end{equation*}
Using the differences between the conditional entropies
\begin{equation}
\delta h_m(\eps):=h_{m-1}(\eps)-h_m(\eps)
\label{eq:deltah}
\end{equation}
the excess entropy can be rewritten as
\begin{equation}
E_m(\eps)=\sum_{k=1}^{m-1} k \delta h_k(\eps) \;.
\label{eqn:Em_deltah}
\end{equation}
Note that the difference $\delta h_m(\eps)$ is the conditional mutual information
\[
\delta h_m(\eps)=I(\vec X_t:\vec X_{t-m}|\vec X_{t-1},\ldots,\vec X_{t-m+1}) \; .
\]
It measures dependencies over $m$ time steps that are not captured by dependencies over $m-1$ time steps. In other words, how much uncertainty about $X_t$ can by reduced
 if in addition to the $m-1$ step past also the
 $m$\,th is taken into account.
For a Markov process of order $m$ the $\delta h_k$ vanish for $k>m$. In this case the sum \eqn{eqn:Em_deltah} contains only a finite number of terms. On the other side truncating the sum at finite $m$ could be interpreted as approximating the excess entropy by the excess entropy of an approximating Markov process.
What can be said about the scale dependency of the $\delta h_m(\eps)$? From Eqs.~(\ref{eps-entr-m}) and (\ref{eps-entr-D}) follows that $h_m(\eps) \propto -log(\eps)$ for $m \le D$ and $h_m(\eps) \propto \mbox{const}$ for $m \ge D$. Using this and Eq.~(\ref{eq:deltah}) we have to distinguish four cases for deterministic systems. Note that $\delta D=D-\lfloor D \rfloor$ denotes the fractal part of the attractor dimension.
\begin{equation}
\delta h_m(\eps) \propto \left\{  \begin{array}{ccl}
\const & \mbox{if} & m \le D - 1 \\
- (1- \delta D) \log \eps & \mbox{if} & m=\lfloor D \rfloor \\
- \delta D \log \eps &\mbox{if} & m=\lfloor D \rfloor+1 \\
\const & \mbox{if} & m > D+1 \;.
\end{array}\right.
\label{eqn:deltaHCases}
\end{equation}
Thus the sum Eq.~(\ref{eqn:Em_deltah}) can be decomposed into three parts:
\begin{equation}
E_m(\eps)=\sum_{k=1}^{\lfloor D \rfloor -1} k \delta h_k(\eps) + \sum_{k=\lfloor D \rfloor}^{\lfloor D \rfloor+1} k \delta h_k(\eps) + \sum_{\lfloor D \rfloor +2}^{m-1} k \delta h_k(\eps) \;.
\label{eq:E_decomp}
\end{equation}
with the $\eps$ dependence showing up only in the middle term (MT):
\begin{equation}
\textrm{MT}=\sum_{k=\lfloor D \rfloor}^{\lfloor D \rfloor+1} k \delta h_k(\eps) \propto c - \log \frac{\eps}{\eps_D}
\label{eq:E_middle}
\end{equation}
with $\eps_D$ denoting the length scale where the deterministic scaling range starts.
Therefore we have decomposed the excess entropy in three terms: Two ideally $\eps$ independent terms and one $\eps$-dependent term. The first term
\begin{equation}
\EEState=\sum_{k=1}^{\lfloor D \rfloor -1} k \delta h_k + c \label{eqn:EEState}
\end{equation}
will be called ``state complexity'' in the following  because it is related to the information encoded in the state of the system. The constant $c$ was added here in order to ensure that the $\epsilon$-dependent term vanishes at $\epsilon_D$ --- the beginning of the deterministic scaling range. The second $\eps$-independent term
\begin{equation}
\EEMem=\sum_{\lfloor D \rfloor +2}^{m-1} k \delta h_k(\eps) \label{eqn:EEMem}
\end{equation}
will be called ``memory complexity'' because it is related to the dependencies between the states on different time steps.
What we call ``state'' in this context is related to the minimal embedding dimension to see the deterministic character of the dynamics which is $\lfloor D \rfloor+1$ \cite{Ding93}. In order to be able to get a one to one reconstruction of the attractor a higher embedding dimension might be necessary \cite{Sauer91a}.
Both \eps-independent terms together we will call ``core complexity''
\begin{equation}
\EECore=\EEState + \EEMem\,.\label{eqn:EECore}
\end{equation}

So far we only addressed the case of a deterministic scaling range.
In the case of a noisy chaotic system we have to distinguish two $\eps$ regions: the deterministic scaling range described above and the noisy scaling range with $h_m(\eps) \approx h_m^c - d \log \eps$ with $h_m^c$ determined by the noise level. In the stochastic scaling range all $\delta h_m$ become $\eps$-independent and the decomposition (\ref{eq:E_decomp}) seems to become unnecessary. This is not completely the case. Let us assume that the crossover between the two regions happens at a length scale $\eps^\ast$. Moreover, let us assume that for sufficiently large $m$ we have in the deterministic scaling range $h_m(\eps) \approx h_{KS}$ (cf. Figs.~\ref{fig:lorenz:quant}\ref{l:0:h2}--\ref{l:n2:h2}). Then we have
\begin{equation}
h_m(\eps^{\ast}) \approx h_{KS} \approx h_m^c- \log \eps^{\ast}\label{eqn:eps-star}
\end{equation}
which allows to express the cross-over scale $\eps^\ast$ in terms of the KS-entropy and the noise level related continuous entropy $h_m^c$
\begin{equation}
\log \eps^{\ast} \approx h_m^c -h_{KS} \;.
\label{eq:eps_ast}
\end{equation}
Moreover, the excess entropy in the \emph{deterministic scaling range} will behave as
\begin{equation}
E_m(\eps) \approx E_{state}+E_{mem} -D \log \frac{\eps}{\eps_{D}} \qquad \text{for} \quad \eps \ge \eps^\ast\;.
\label{Em-det-scaling}
\end{equation}
Evaluating this expressing at the crossover length scale $\eps^\ast$ allows to express the value of the excess entropy in the \emph{stochastic scaling range} as
\begin{equation}
E_m \approx E_{state}+E_{mem} + D \log \eps_{D} - D (h_m^c-h_{KS})\qquad \text{for} \quad \eps \le \eps^\ast\;.
\label{Em-stoch-scaling}
\end{equation}
In particular, this expression shows that in decreasing the noise level, \ie $h_m^c$ will increase the asymptotic value of the excess entropy for noisy systems.
Thus, an increased excess entropy or predictive information for a fixed length scale or partition can be achieved in many ways:
\begin{enumerate}
\item by increasing dimension $D$ of the dynamics
\item by decreasing the noise level $\eps^\ast$
\item by increasing the amplitude $\eps_{D}$
\item by increasing the state complexity
\item by increasing the correlations measured by the ``memory'' complexity, i.e. by increasing the predictability
\item by decreasing the entropy rate $h_{KS}$, i.e. by decreasing the unpredictability
\end{enumerate}

Naturally, the effect of the noise level will be observed in the stochastic scaling range only. In practice there might be more than one deterministic and stochastic scaling range or even no clear scaling range at all. How we will deal with these cases will be described below when we introduce our algorithm.

\subsection{Methods for estimating the information theoretic measures}\label{subsec:methods-estimation}

Reliably estimating entropies and mutual information is very difficult in high-dimensional spaces due to the increasing bias of entropy estimates. Therefore we will employ two different approaches. On the one hand we will use an algorithm for the estimation of the mutual information proposed by \citet{KraskovSG04:EstimatingMI} based on nearest neighbor statistics which allows to reduce the bias by employing partitions of different sizes in spaces of different dimensions. On the other hand we calculate a proxy for the excess entropy using correlation entropies \citep{Takens1998} of order $q=2$. These are related to the R\'enyi entropies of second order and the correlation sum provides an unbiased estimator.
Both methods do not require binning but differ substantially in what they compute.

\paragraph{Estimation via local densities from nearest neighbor statistics (KSG)}

The most common approach to estimate information quantities of continuous processes, such as the mutual information, is to
 calculate the differential entropies \eqnp{eqn:entropy:diff}
 directly from  the nearest neighbor statistics.
The key idea is to use nearest neighbor distances \citep{Vasicek1976,Dobrushin1958,Kozachenko1987}
as proxies for the local probability density.
This method corresponds in a way to an adaptive bin-size for each data point.
For the mutual information $I(X,Y)$ (required \eg to calculate the PI \eqnp{eqn:PI}), however, it is not recommended to
 naively calculate it directly from the individual entropies of $X$, $Y$ and their joint $(X,Y)$
 because they may have very dissimilar scale
 such that the adaptive binning leads to spurious results.
For that reason a new methods was proposed by \citet{KraskovSG04:EstimatingMI},
 that we call \emph{KSG},
 which only uses the nearest neighbor statistics in the joint space.
We denote $\IKSG{k}(X,Y)$ the mutual information estimate where $k$ nearest neighbors where used for the local estimation.

The length scale on which the mutual information is estimated by this algorithm
 depends on the available data.
In the limit of infinite amount of data $\IKSG{k}(X,Y) = \lim_{\eps \to 0} I(X,Y,\eps)$ for $k\gg 1$.
However, in order to evaluate the quantity at a certain length scale (similar to $\eps$ above) and
 assuming the same underlying space for $X$ and $Y$,
 noise of strength $\eta$ is added to the data resulting in
\begin{align}
  \IKSG{k}(X,Y,\eta) = \IKSG{k}(X+U(\eta),Y+U(\eta))\label{eqn:I:KSG}
\end{align}
where $U(\eta)$ is the uniform distribution in the interval $[0,\eta]$.
The idea of adding noise is to make the processes X and Y independent within
 neighborhood sizes below the length scale of the noise.
In this way only the structures above the added noise-level contribute to the mutual information.
Note that for small $\eta$ the actual scale ($k$-neighborhood size)
 may be larger due to sparsity of the available data.

\paragraph{Estimation via correlation sum}

The correlation sum is one of the standard tools in nonlinear time series analysis~\cite[Chapter~6]{kantz2004nonlinear}, \cite{GPA@scholarpedia}. Normally it is used to estimate the attractor dimension. However, it can also be used to provide approximate estimates of entropies and derived quantities such as the excess entropy. The correlation entropies for a random variable $\vec X$ with measure $\mu(\vec x)$ are defined as \citep{Takens1998}

\begin{equation}
H^{(q)}(\vec{X};\eps) = \left\{ \begin{array}{cc}

- \frac{1}{q-1}  \log \int \mu(B(\vec x,\eps))^{q-1} d\mu(\vec x) & \mbox{for} \quad q \neq 1 \\
- \int \log \mu(B(\vec x,\eps)) d \mu(\vec x) & \mbox{for} \quad  q=1
\end{array}
\right.
\label{eq:Hq}
\end{equation}
where $B(\vec x,\eps)=\{\vec y \; | \;||\vec y - \vec x|| < \eps \}$ is the ``ball'' at $\vec x$ with radius $\eps$.
For $q=2$ \eqnp{eq:Hq} becomes $H^{(2)}(\vec{X};\eps) = -\log \int \mu(B(\vec x,\eps)) d\mu(\vec x)$. The integral in this formula is also known as ``correlation integral''.
For $N$ data points $\vec x_i \in\Real^n \quad i=1,\ldots,N$ it can be estimated using the correlation sum, which is the averaged relative number of pairs in an  $\eps$-neighborhood~\cite[Chapter~6]{kantz2004nonlinear},\cite{GPA@scholarpedia}

\begin{equation}
    C^{(2)}(\eps) = \frac{2}{N (N-1)}\sum_{i=1}^{N-1} \sum_{j=i+1}^N \Theta\left(\eps - \|\vec x_i-\vec x_j\|\right) \;.
\end{equation}
$\Theta$ denotes the Heaviside function
\[
\Theta(x)=\left\{ \begin{array}{ccc}
0 & : & x < 1 \\
1 & : & x \ge 1\,.
\end{array}  \right.
\]
Then the correlation entropy is
\begin{align*}
  H^{(2)}(\eps) &= -\log C^{(2)}(\eps) \;.
\end{align*}
For sufficiently small $\eps$ it behaves as
\begin{align*}
  H^{(2)}(\eps) &\approx \const -D^{(2)} \log \eps
\end{align*}
with $D^{(2)}$ being the correlation dimension of the system \citep{Grassberger1983:StrangeAttractors}. A scale dependent correlation dimension can be defined as difference quotient

\begin{equation}
D^{(2)}(\eps)=\log \frac{C^{(2)}(\eps+\Delta)}{C^{(2)}(\eps)} \left( \log \frac{\eps+\Delta}{\eps} \right)^{-1} \;.
\label{eqn:D2}
\end{equation}

For a temporal sequence of states (or state vectors) $x_i \quad i=1,\ldots,m$ we can now define block entropies $H_m^{(2)}(\eps)= -\log C^{(2)}(\eps)$ by using $m \times d$-dimensional delay vectors $\vec x_i$. Now, we can define also the quantities corresponding to conditional entropies and to the excess entropy using the correlation entropy
\begin{align}
h_m^{(2)}(\eps)&=H_{m+1}^{(2)}(\eps)-H_{m}^{(2)}(\eps) \label{eqn:h2m} \\
E_m^{(2)}(\eps) &= H_{m}^{(2)}(\eps) - m h^{(2)}_{m-1}(\eps) \;. \label{eqn:E2m}
\end{align}

We expect the same qualitative behavior regarding the $\eps$-dependence of these quantities as for those based on Shannon entropies, see \eqns{eqn:hn}{eqn:Em}. Quantitatively there might be differences, in particular for large $\eps$ and strongly non-uniform measures.

A comparison of the two methods with analytical results are given in the Appendix \ref{sec:AR2}, where we
 find a good agreement. Although, the KSG method seems to underestimate the mutual
information for larger embeddings (higher dimensional state space).
The correlation integral method uses a unit ball of diameter $2\eps$ whereas
 the KSG method measures the size of the hypercube enclosing $k$ neighbors
  where the data was subject to additive noise in the interval $[0,\eta]$.
Thus comparable results are obtained with $\eta \approx 2 \eps $.

\subsection{Algorithm for excess entropy decomposition}
\label{sec:decomp:algo}

We are now describing the algorithm used to compute the proposed decomposition of the excess entropy in Sec.~\ref{sec:decomp}.
The algorithm is composed of several steps: preprocessing, determining the middle term (MT) \eqnp{eq:E_middle}, determining the constant in MT, and
 the calculation of the decomposition and of quality measures.

\paragraph{Preprocessing:} Ideally the $\delta h$ curves are composed of straight lines in a log-linear representation, \ie of the form
 $o - s \log(\eps)$. We will refer to $s$ as the slope (it is actually the inverted slope). Thus we perform fitting, that attempts to find segments following this form, details are provided in the Appendix~\ref{sec:app:decomp}.
Then the data is substituted by the fits in the intervals where the fits are valid.
As for very small scales the $\delta h$ become very noisy we extrapolate below the fit with the smallest scale.
In addition we calculate the derivative $\hat s(m,\eps)=\frac{d\,\delta h_m}{d\,\eps}$ in each point, either from the fits ($s$, where available) or from finite differences of the data (using 5 points averaging).

\paragraph{Determining MT:}
In theory only two $\delta h_m$ should have a non-zero slope at each scale $\eps$, see \eqn{eqn:deltaHCases}.
However, in practice we have often more terms, such that we
 need to find for each $\eps$ the maximal range $(m_l,m_u)$,
 where $\forall i\in [m_l,m_u]:  \hat s(i,\eps)>s_{\textrm{min}}$, \ie the slope is larger than the threshold $s_{\textrm{min}}$.
However, this is only valid for deterministic scaling ranges. In stochastic ranges all $\delta h$ should have zero slope.
We introduce a measure of stochasticity, defined as $\kappa(m,\eps) = 1-\sum_{k=m_l}^{m_u} \hat s(k,\eps)$ which is 0 for purely deterministic ranges and 1 for stochastic ones.
The separation between state and memory complexity is then inherited from the next larger deterministic range. Thus if $\kappa(m,\eps)\ge\kappa_{\textrm{max}}$ we use $(m_l,m_u)$ at $\eps^*$, where $\eps^* =\argmin_{e \in (\eps, \infty)} \kappa(m,e)<\kappa_{\textrm{max}}$. Note that the here algorithmically defined $\epsilon^*$ is not necessarily equal to the $\epsilon^{\ast}$ defined above (\ref{eq:eps_ast}) for an ideal-typical noisy deterministic system.

\paragraph{Determining the constant in MT:}
In order to obtain the scale-invariant constant $c$ of the MT, see \eqn{eq:E_middle}, we would have to define
 a certain length scale $\eps_D$. Since this cannot be done robustly in practice (in particular because it may not be the same $\eps_D$ for each $m$) we resort to a different approach.
The constant parts of the $\delta h_{m_l\ldots m_u}$ terms in the MC can be determined from plateaus on larger scales.
Thus, we define $c_m^{\textrm{MT}}(\eps)=\min_{e \in (\eps,\eps^*]}\delta h_m(e)$, where $\eps^*$ is smallest scale $>\eps$ where we have a near-zero slope, \ie. $\eps^* =\argmin_{e \in (\eps,\infty)} s(m,e)<s_{\textrm{min}}$. In case there is no such $\eps^*$ then $c_m^{\textrm{MT}}=0$.

\paragraph{Decomposition and quality measures:}
The decomposition of the excess entropy follows \eqns{eqn:EEState}{eqn:EEMem} with $m_l$ and $m_u$ used for splitting the terms:
\begin{align}
  \EEState&=\sum_{k=1}^{m_l-1} k \delta h_k  + \sum_{k=m_l}^{m_u} k c_k^{\textrm{MT}}\qquad\EEMiddle=\sum_{k=m_l}^{m_u} k (\delta h_k - c_k^{\textrm{MT}})\qquad\EEMem=\sum_{k=m_u+1}^{m} k \delta h_k \label{eqn:E_decomp:algo}
\end{align}
In addition we can compute different quality measures to indicate the reliability of the results, see \tab{tab:quality}.

\begin{table}
  \centering
  \begin{tabular}{r|l|l}
    quantity & definition & description\\\hline
    $\kappa$=stochastic(\eps) & $1-\sum_{k=m_l}^{m_u} \hat s(k,\eps)$ &0: fully deterministic, 1: fully stochastic at $\eps$\\
    \% negative(\eps) & $1/m\sum_{k=1}^{m} \llbracket \delta h(k,\eps)<0 \rrbracket$ &percentage of negative $\delta h$ \\
    \% no fits(\eps) & $1/m\sum_{k=1}^{m} \llbracket \textrm{valid fit for } \delta h(k,\eps) \rrbracket$ &percentage of $\delta h$ where no fit is available\\
    \% extrap.(\eps) & $1/m\sum_{k=1}^{m} \llbracket \textrm{is } \delta h(k,\eps) \textrm{ extrapolated}  \rrbracket$ &percentage of $\delta h$ that where extrapolated\\
  \end{tabular}
  \caption{Quality measures for decomposition algorithm. We use the Iverson bracket for Boolean expression: $\llbracket \textrm{True} \rrbracket:=1$ and $\llbracket \textrm{False} \rrbracket:=0$. They are all normalized to $[0,1]$ where typically 0 is the best score and 1 is the worst.}\label{tab:quality}
\end{table}

\subsection{Illustrative example}\label{ss:Example}

To get an understanding of the quantities, let us first apply the methods
 to the Lorenz attractor, as a deterministic chaotic system,
 and to its noisy version as a stochastic system.

\begin{figure}
  \centering
  \begin{tabular}{cc}
    (a) phase-space $(x,y,z)$ & (b) delay embedding of $x$\\
    \includegraphics[height=.4\columnwidth]{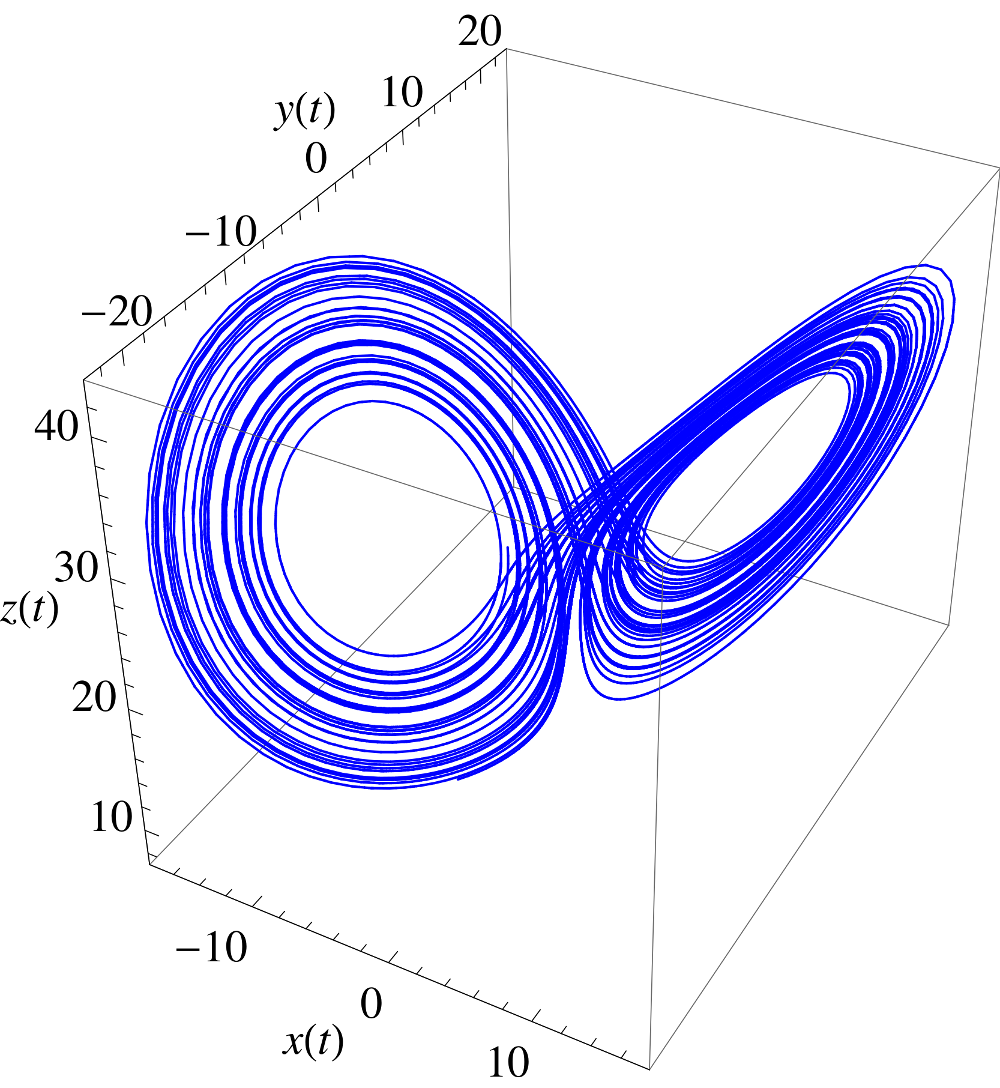}&
    \includegraphics[height=.4\columnwidth]{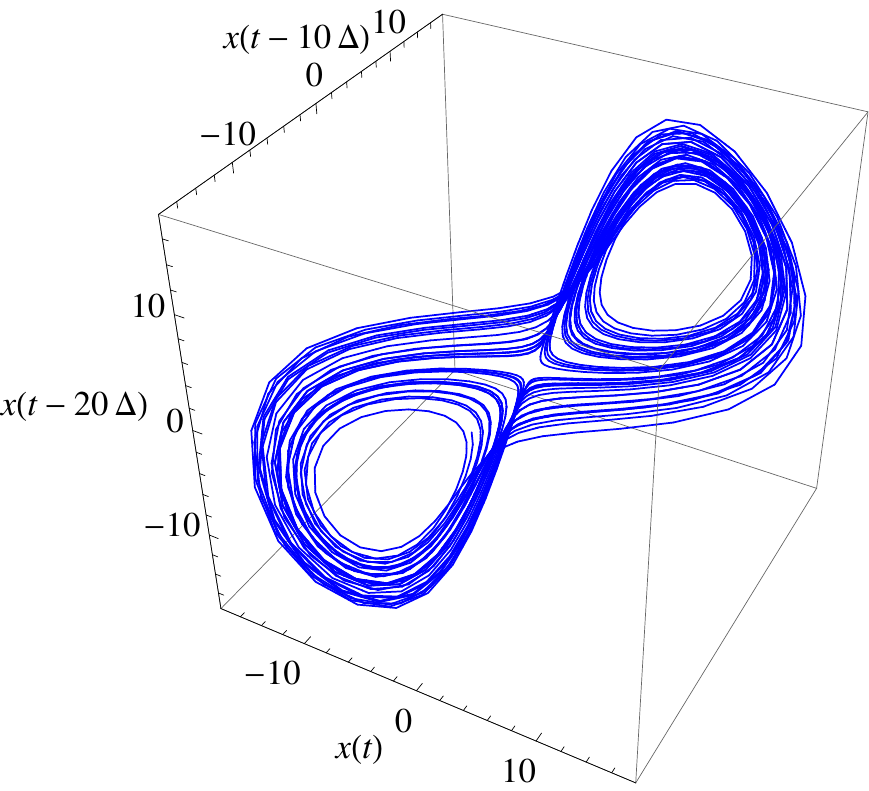}
  \end{tabular}
  \caption{Lorenz attractor with standard parameters.
    \infig{(a)}~Trajectory in state space $(x,y,z)$.
    \infig{(b)}~Embedding of $x$ with $m=3$, time-delay $\tau=10$, and $\Delta=0.01$.
  }
\label{fig:lorenz}
\end{figure}

\paragraph{Deterministic system: Lorenz attractor}
The Lorenz attractor is obtained as the solution to the following differential equations:
\begin{align}
  \dot x &= s(-x+y)     \label{eqn:lorenz:x} \\
  \dot y &= -xz +rx - y \label{eqn:lorenz:y}\\
  \dot z &= -xy -bz     \label{eqn:lorenz:z}
\end{align}
integrated with standard parameters $s=10$, $r=28$, $b=8/3$.
\Fig{fig:lorenz} displays the trajectory in phase-space ($x,y,z$)
 and using delay embedding of $x: (x(t), x(t-\tau\Delta), \dots , x(t-m\tau\Delta))$ using $m=3$
 and time delay is $\tau=10$ (for sampling time $\Delta=0.01$).
The equations where integrated with Runge-Kutta fourth order with step size $5\cdot 10^{-4}$.
An alternative time delay may be obtained from first minimum of mutual information
 as a standard choice~\citep{Fraser86, Olbr97b}, which would be $\tau=18$, but
$\tau=10$ yields clearer results for presentation.

\begin{figure}
  \centering
  \begin{tabular}{ccc@{\hspace{-10pt}}l}
    Deterministic & Noise 0.005 & Noise 0.01\\
    \subplot{l:0:d2} & \subplot{l:n1:d2} & \subplot{l:n2:d2}\\[-1.1em]
    \includegraphics[width=.3\columnwidth]{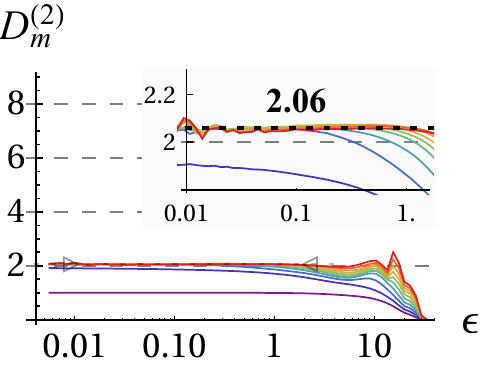}&
    \includegraphics[width=.3\columnwidth]{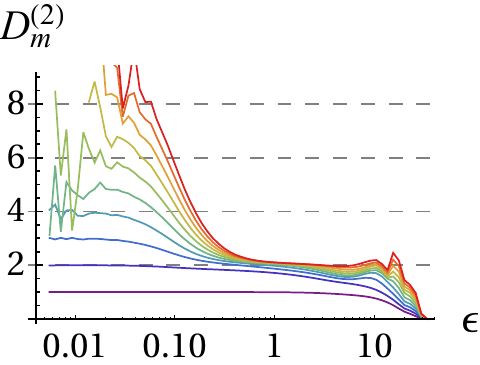}&
    \includegraphics[width=.3\columnwidth]{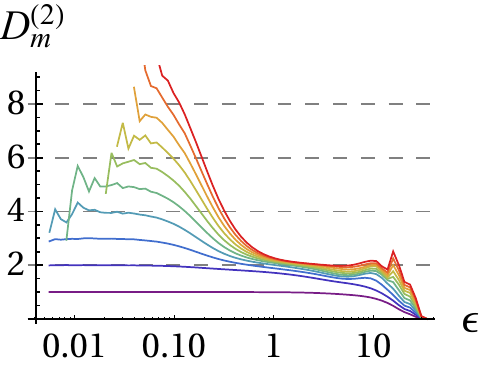}&
    \multirow{2}{*}{\includegraphics[scale=1,trim=0 0 0 -10]{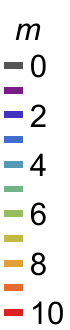}}\\
    \subplot{l:0:h2} & \subplot{l:n1:h2} & \subplot{l:n2:h2}\\[-1.1em]
    \includegraphics[width=.3\columnwidth]{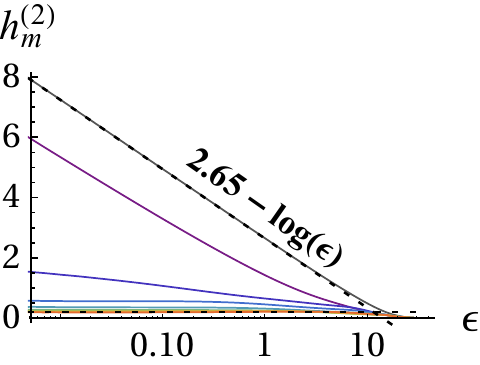}&
    \includegraphics[width=.3\columnwidth]{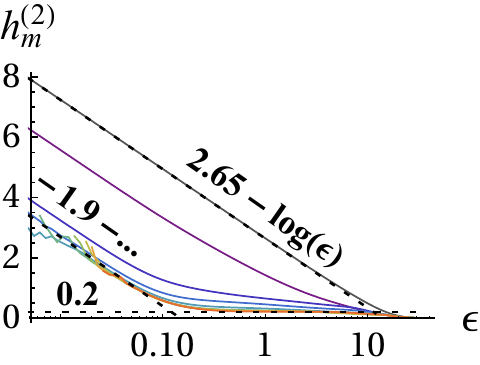}&
    \includegraphics[width=.3\columnwidth]{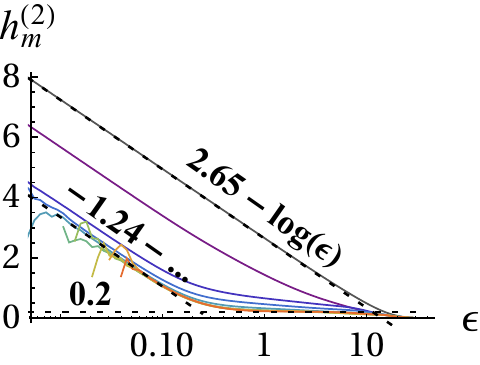}\\
    \subplot{l:0:e2} & \subplot{l:n1:e2} & \subplot{l:n2:e2}\\[-1.1em]
    \includegraphics[width=.3\columnwidth]{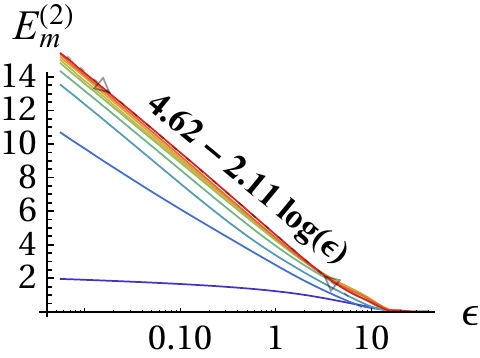}&
    \includegraphics[width=.3\columnwidth]{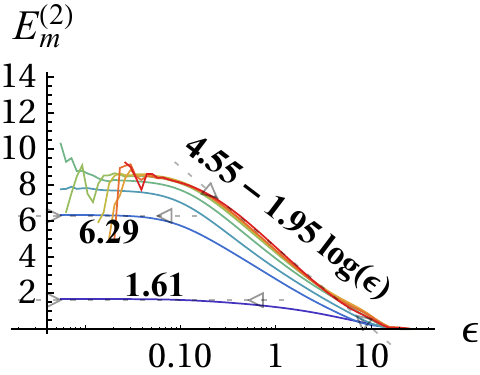}&
    \includegraphics[width=.3\columnwidth]{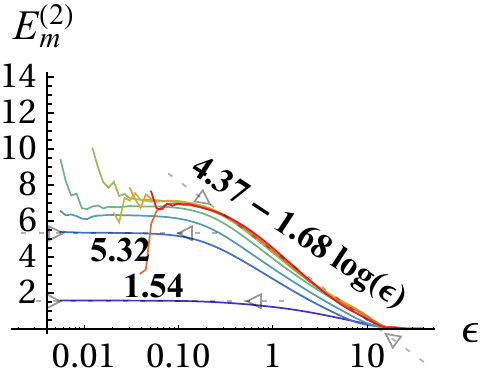}\\
    \subplot{l:0:pi} & \subplot{l:n1:pi} & \subplot{l:n2:pi}\\[-1.1em]
    \includegraphics[width=.3\columnwidth]{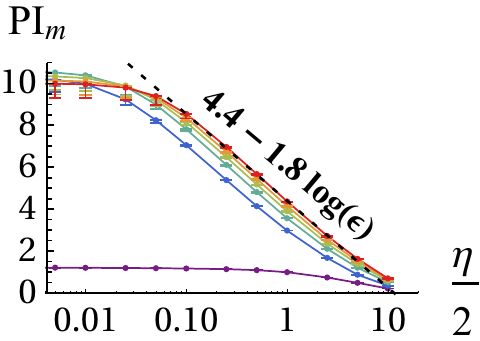}&
    \includegraphics[width=.3\columnwidth]{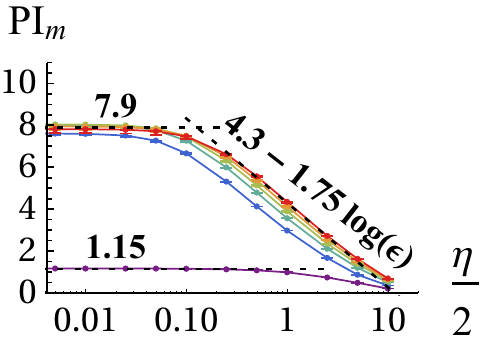}&
    \includegraphics[width=.3\columnwidth]{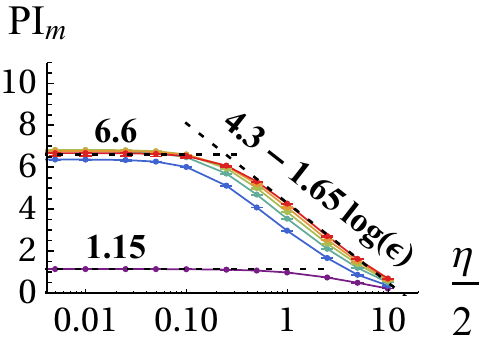}&
    \includegraphics[scale=1,trim=0 -35 0 0]{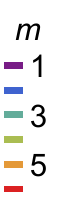}
  \end{tabular}\\
  \caption{Correlation dimension $\Dtwo_m$~\infig{\ref{l:0:d2}--\ref{l:n2:d2}} \eqnp{eqn:D2},
    conditional block entropies $\htwo_m$~\infig{\ref{l:0:h2}--\ref{l:n2:h2}} \eqnp{eqn:h2m}, %
    excess entropy $\Etwo_m$ ~\infig{\ref{l:0:e2}--\ref{l:n2:e2}} \eqnp{eqn:E2m}, and
    predictive information $\PI_m$ ~\infig{\ref{l:0:pi}--\ref{l:n2:pi}} \eqnp{eqn:PI}
    of the Lorenz attractor estimated with
    the correlation sum method \ref{l:0:d2}--\ref{l:n2:e2} and KSG \ref{l:0:pi}--\ref{l:n2:pi}
    for the deterministic system (first column)
    and with absolute dynamic noise $0.005$, and $0.01$ (second and third column respectively).
    The error bars in \ref{l:0:pi}--\ref{l:n2:pi} show the values calculated on half of the data.
    All quantities are given in nats (natural unit of information with base $e$) and in
    dependence of the scale ($\eps$,$\eta$ in space of $x$ \eqn{eqn:lorenz:x}) for a
    range of embeddings $m$, see color code.
    In \ref{l:0:h2}--\ref{l:n2:h2} the fits for $h_0 = H_1$ allow to determine $H^C\!= 2.65$ %
    and for $h_{m}$ give $h_{KS}$ and $h^c_{m}$, \eqnr{eqn:eps-star}{Em-stoch-scaling}.
     Parameters: delay embedding of $x$ with $\tau=10$.
  }
\label{fig:lorenz:quant}
\end{figure}

The result of the correlation sum method using TISEAN~\cite{tisean1999} is depicted in \fig{fig:lorenz:quant} for $10^6$ data points. From the literature we know that the chaotic attractor has a fractal dimension of $2.06$~\cite{Grassberger1983:StrangeAttractors} which we can verify with the correlation dimension $\Dtwo_m$, see \fig{fig:lorenz:quant}\ref{l:0:d2} for $m>2$ and small $\eps$.
The conditional $\htwo_m$ \ref{l:0:h2} becomes constant and identical for larger $m$.
The excess entropy $\Etwo_m$ \ref{l:0:e2} approaches the predicted scaling behavior of:
$\const-D\log(\eps)$ \eqnp{eqn:Em-det} with $D=2.06$ and $\const=4.75$.
The predictive information $\PI_m$ \ref{l:0:pi} shows the same scaling behavior on the coarse range
 but with a smaller slope and saturates at small \eps.

\resetsubplot

\resetsubplot
\begin{figure}
  \centering
  \begin{tabular}{c@{\hspace{-3pt}}c@{\hspace{-3pt}}c}
    \subplot{l:0:de} Deterministic & \subplot{l:n1:de} Noise 0.005 & \subplot{l:n2:de} Noise 0.01\\
    \includegraphics[width=.33\columnwidth]{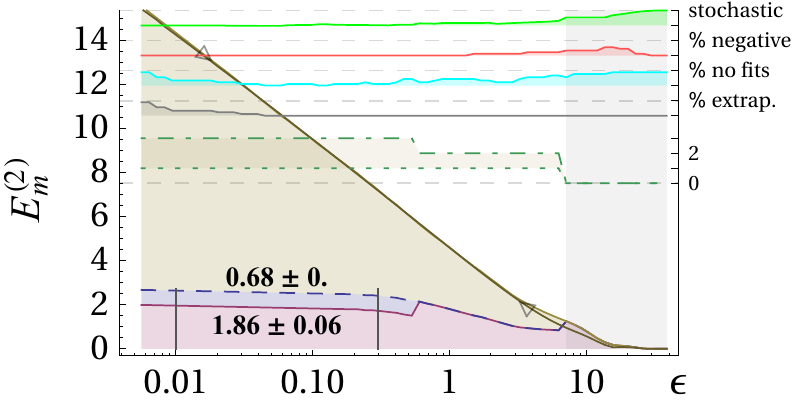}&
    \includegraphics[width=.33\columnwidth]{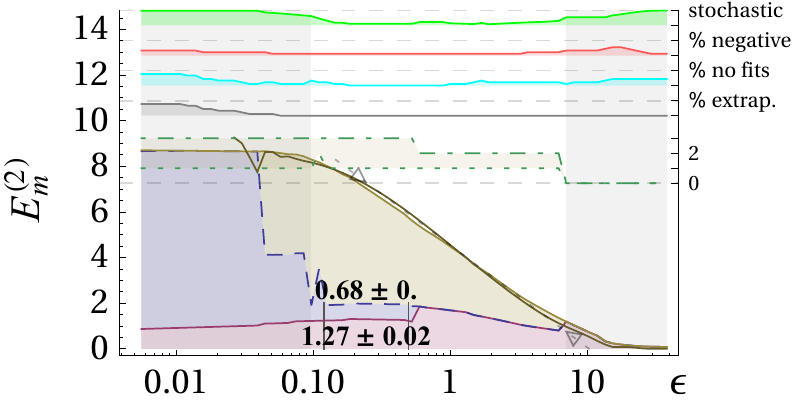}&
    \includegraphics[width=.33\columnwidth]{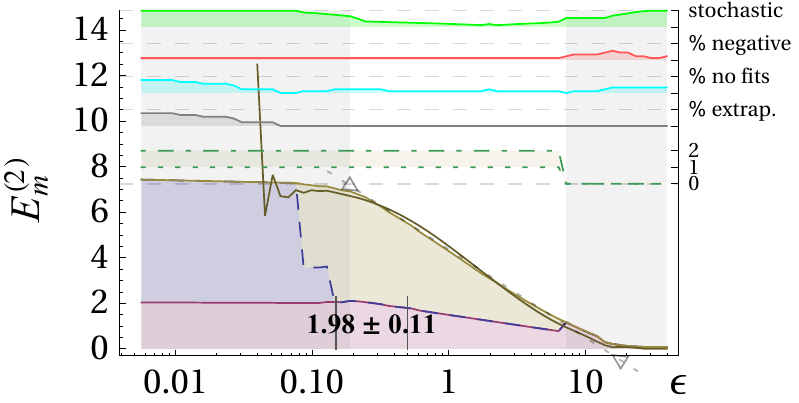}\\
  \end{tabular}
  \includegraphics[width=.95\columnwidth]{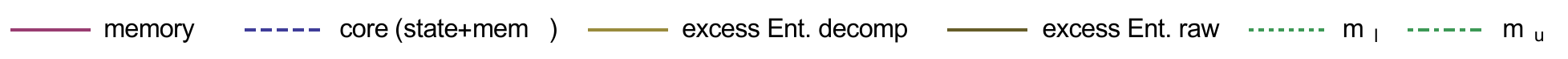}
  \caption{Excess Entropy decomposition for the Lorenz attractor into state complexity (blue shading), memory complexity (red shading), and \eps-dependent part $E_{\eps}$ (beige shading) (all in nats).
    Columns as in \fig{fig:lorenz:quant}.
    A set of quality measures and additional information is displayed on the top using the right axis.
    $m_l$ and $m_u$ refer to the terms in $E_{\eps}$ \eqnp{eqn:E_decomp:algo}.
    Scaling ranges that are identified as stochastic are shaded in gray (stochastic indicator $>\kappa_{\textrm{max}}=0.5$).
    In manually chosen ranges, marked with vertical black lines, we evaluate the mean and standard deviation of the memory- and state complexity.
    Parameters: $s_{\textrm{min}}=0.1$.
  }
  \label{fig:lorenz:decomp}
\end{figure}

\paragraph{Stochastic system: noisy Lorenz attractor}
In order to illustrate the fundamental differences between deterministic and stochastic systems
 when analyzed with information theoretic quantities we consider now the Lorenz dynamical system
 with dynamic noise (additive noise to the state ($x,y,z$) in \eqnr{eqn:lorenz:x}{eqn:lorenz:z} before each integration step) as provided by the TISEAN package\cite{tisean1999}.
The dimension of a stochastic systems is infinite, \ie
for embedding $m$ the correlation integral yields the full embedding dimension as shown
 in \fig{fig:lorenz:quant}\ref{l:n1:d2},\ref{l:n2:d2} for small \eps.

On large scales the dimensionality of the underlying deterministic system can be revealed to
 some precision either from $\Dtwo$ or from the slope of $\Etwo$ or $\PI$.
Thus, we can identify a determinstic and a stochastic scaling range with different qualitative scaling behavior of the quantities.
In contrast to the deterministic case, the excess entropy converges in the stochastic scaling range
 and no more contribution from $m>5$ are observed as shown in \fig{fig:lorenz:quant}\ref{l:n1:e2},\ref{l:n2:e2}.
By measuring the \eps-dependence (slope)
 we can actually differentiate between deterministic and stochastic systems and scaling ranges, which
 is performed by the algorithm (\sec{sec:decomp:algo}) and presented in \fig{fig:lorenz:decomp}.
The predictive information \ref{l:n1:pi}--\ref{l:n2:pi} again yields a lower slope, meaning a lower dimension estimate,
but is otherwise consistent.
However, it does not allow to safely distinguish determinstic and stochastic systems,
 because it always saturates due to the  effect of finite amount of data.

Let us now consider the decomposition of the excess entropy as discussed in \sec{sec:decomp} and \sec{sec:decomp:algo} to see whether the method is consistent.
\Fig{fig:lorenz:decomp} shows the scale-dependent decomposition together with the determined stochastic and deterministic
 scaling ranges.
The resulting values for comparing the determinstic scaling range are given in \tab{tab:lorenz:constants} and reveal, that the
$\EECore$ decreases for increasing noise level similarly as the constant reduces.
This is a consistency check because the intrinsic scale (amplitudes) of the deterministic and noisy Lorenz system are almost identical. We also checked the decomposition for a scaled version of the data where we found the same values state-, memory- and core complexity.
We see also that the state complexity $\EEState$ vanishes for larger noise, because the dimension drops significantly below 2, where we have no
summands in the state complexity and the constant $c^{\textrm{MT}}$ is also 0.

\begin{table}
  \centering
  \begin{tabular}{ll|cccc}
                       &                       & determ.       &noise $0.005$&noise $0.01$    &noise $0.02$\\
    \hline
    \eqn{eqn:Em-det}       & $D$               & 2.11          & 1.95          & 1.68         & 1.59 \\
    \eqn{eqn:Em-det}       & $\const$          & 4.62          & 4.55          & 4.37         & 4.21 \\
    \eqn{eqn:eps-star}     & $\eps^{\ast}$     & 0              & 0.12          & 0.23       & 0.45 \\
    \eqn{eqn:EEState}      & \EEState          & 0.68$\pm$0    & 0.68$\pm$0    & 0            & 0    \\
    \eqn{eqn:EEMem}        & \EEMem            & 1.86$\pm$0.06 & 1.27$\pm$0.02 & 1.98$\pm$0.11 & 1.2$\pm$0.1\\
    \eqn{eqn:EECore}       & \EECore           & 2.54$\pm$0.06 & 1.95$\pm$0.02 & 1.98$\pm$0.11 & 1.2$\pm$0.1\\
  \end{tabular}
  \caption{Decomposition of the excess entropy for the Lorenz system.
    Presented are the unrefined constant and dimension of the excess entropy,
    the state-, memory-, and core complexity (in nats) determined in the deterministic scaling range, see \fig{fig:lorenz:decomp}.
   }\label{tab:lorenz:constants}
\end{table}

To summarize, in deterministic systems with a stochastic component, we can identify different $\eps$-ranges (scaling ranges) within which  stereotypical behaviors of the entropy quantities are found.
This allows us to determine the dimensionality of the deterministic attractor and decompose the excess entropy in order to obtain two \eps-independent complexity measures.

\section{Results}\label{s:Results}

In this section we will apply our methods to the behavior of robotic systems.
We use a previously published control scheme that can generate a large variety of smooth and
coherent behaviors from a local learning rule~\cite{DerGSO2012, DerMartius13}.
As it is a simplified version of the time-local predictive information maximization controller
 we aim at answering the hypothesis that the controller actually
 increases the complexity of behavior during learning.

\subsection{Application to robotics: controller and learning} \label{ss:RobotControlLearn}

Let us now briefly introduce the control framework and the robots used to generate
 the behavior we are analyzing in the following.
The robots are simulated in a physically realistic rigid-body simulation~\cite{lpzrobots10}. Each robot
 has a number of sensors producing real-valued stream $s\in\Real^n$ and a number of actuators, also
 controlled by real values $a\in\Real^o$.
The sensors measure the actual joint angles and the body accelerations.
 In all examples the actuators are position controlled servo motors with the particular property to have no power around the set-point. This enables small perturbations
 to be perceivable in the joint-position sensor values.
The time is discretized with an update frequency of 50\,Hz.

The control is done in a reactive closed-loop fashion by a one layer feed-forward neural network:
\begin{align}
  a_t=K(s_t,C,h) = \tanh(C s_t + h),\label{eqn:controller}
\end{align}
where $C$ is a weight matrix, $h$ is a weight vector, and $\tanh$ is applied component-wise. For fixed weights this controller is very simple, however it can
 generate non-trivial behavioral modes due to the complexity of the body-environment interaction.
For instance with a scaled rotation matrix $C$ an oscillatory behavior may be controlled
 which frequency is adaptive to how fast the physical system follows.
To get a variety of behavioral modes the parameters ($C,h$) have to be changed which can be done
 by optimizing certain objective functions.
Several methods have been proposed that use generic (robot independent) internal drives.
Homeokinesis~\cite{derspherical06, DerMartius11} for instance balances dynamical instability with predictability. More recently the  maximization of the time-local predictive information of the sensor process
 was used as driving force~\cite{MartiusDerAy2013}.
On an intuitive level the maximization of the PI~\eqnp{eqn:PI} of the sensor process leads to a high variety of sensory input due to the entropy term $H(S)$ while keeping a temporal structure by minimizing the future uncertainty reflected in the conditional entropy $H(S'|S)$.
Indeed we have found that it produces smooth and coordinated behavior in different robotic systems, including hexapod robots and humanoids.
On the other hand, maximizing PI also leads to behavior of high complexity with maximal dimension as discussed in \sec{sec:decomp}.
A simplified version of the PI-maximizing algorithm, called ULR published in~\cite{DerGSO2012, DerMartius13}, leads to more coherent and defined behavioral modes of seemingly low dimensionality which we use in this paper.
In order to apply the method to behavior of robots or possibly to that of animals
the state space reconstruction has to be done from certain measurements.
Here we use a single sensor value, but the formalism can be extended to multiple sensor values as well.
It must be guaranteed that sufficient physical coupling exist between the measured quantity and the rest of the  body.

\subsection{Experiments}\label{Experiments}

\begin{figure}
  \centering
  \begin{tabular}{cc}
    (a) \Hexapod{} &(b) \Snake\\
    \includegraphics[width=.45\linewidth]{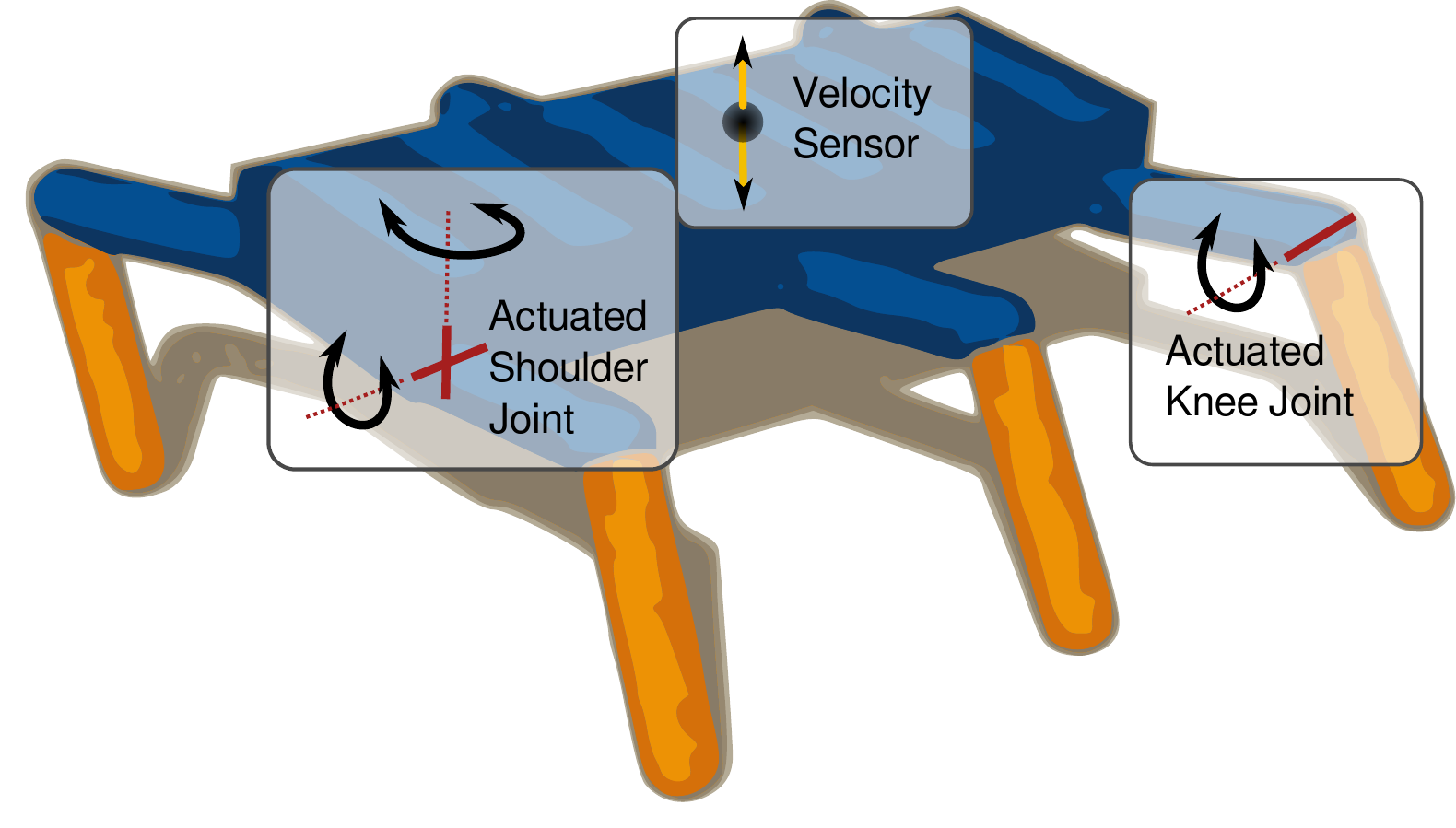}&
    \hspace*{-1cm}\includegraphics[width=.55\linewidth]{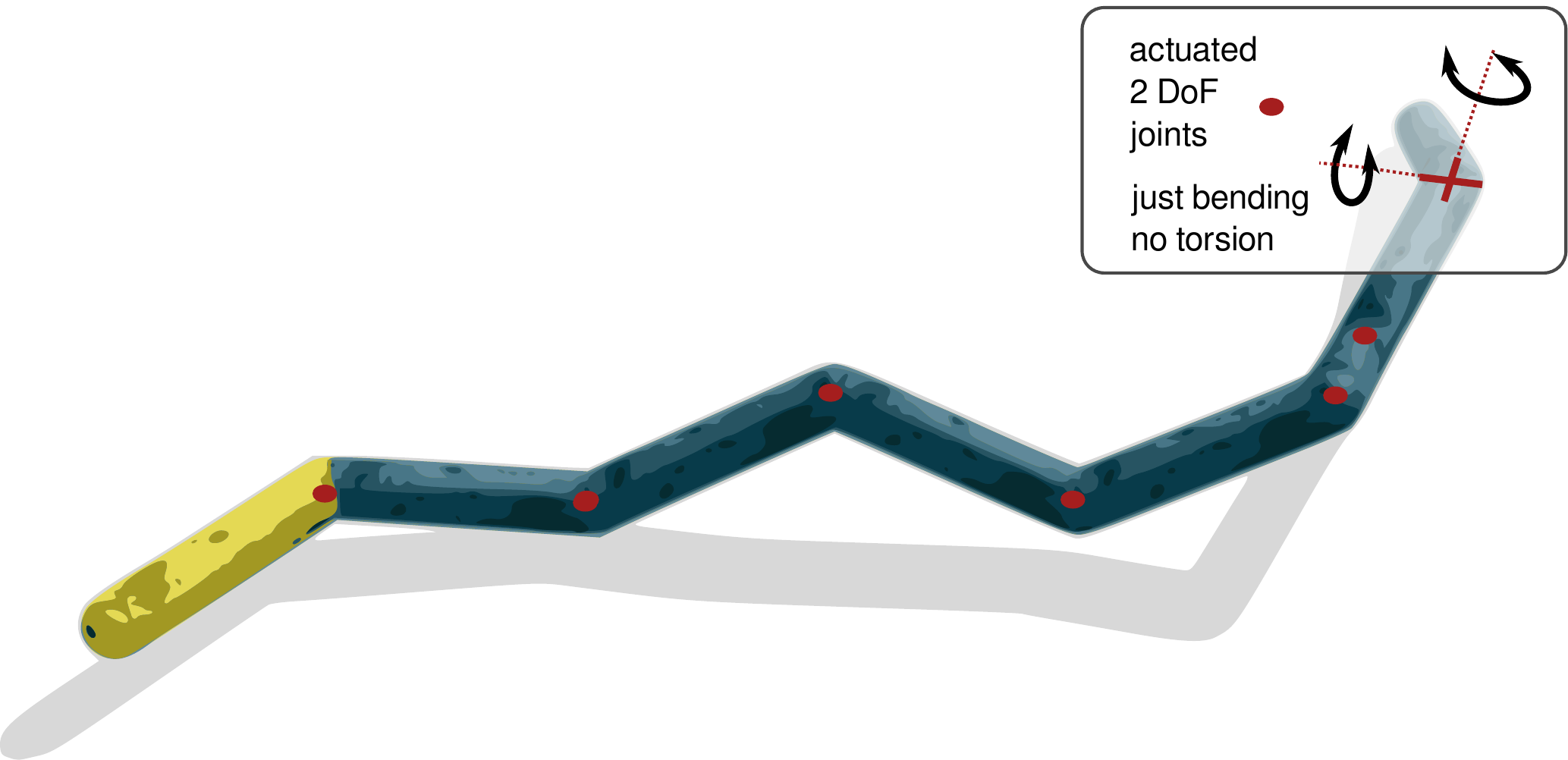}
  \end{tabular}
  \caption{Robots used for the experiments.
    \infig{(a)} The \Hexapod{}. 18 actuated DoF: 2 shoulder and 1 knee joint per leg.
    \infig{(b)} The \Snake{}. 14\,DoF: 8 segments connected by 2\,DoF joints each.}
  \label{fig:robots}
\end{figure}

\begin{table}
  \centering
  \begin{tabular}{|l|l|l|l|l|}
    \hline
    Robot&Behavior&Length&Dataset&Video\\
    \hline
    \Snake&side rolling &$9\cdot 10^5$&D-S1&\href{http://playfulmachines.com/QuantBeh2015/#vid:s1}{Video S1}\\
    \hline
    \Snake&crawling     &$9\cdot 10^5$&D-S2&\href{http://playfulmachines.com/QuantBeh2015/#vid:s2}{Video S2}\\
    \hline
    \Hexapod&jumping after 2\,min&$10^6$&D-H1&\href{http://playfulmachines.com/QuantBeh2015/#vid:h1}{Video H1}\\
    \hline
    \Hexapod&jumping after 4\,min &$10^6$&D-H2&\href{http://playfulmachines.com/QuantBeh2015/#vid:h2}{Video H2}\\
    \hline
    \Hexapod&jumping after 8\,min &$10^6$&D-H3&\href{http://playfulmachines.com/QuantBeh2015/#vid:h3}{Video H3}\\
    \hline
  \end{tabular}
  \caption{Experiments, data sets and videos.
    \href{http://playfulmachines.com/QuantBeh2015/}{See \tt{\small playfulmachines.com/QuantBeh2015}}.
}
  \label{tab:experiments}
\end{table}
Let us now present the emergent behaviors obtained from the control algorithm ULR.
We use two robots, a snake-like robot and an insect-like hexapod robot as illustrated in \fig{fig:robots}.
The behaviors of the snake are published here for the first time, whereas the behaviors of the hexapod
have been published before in \citet{DerMartius13}.
However, this paper focuses primarily on the quantification of behavior and not on its generation.
The \Snake has identical cylindrical segments which are connected by actuated $2$\,degrees of freedom (DoF) universal joints.
In this way two neighboring segments can be inclined to each other, but without relative torsion along the segment-axes.
The sensor values are the actual positions of the joints.
If the robot is controlled with the ULR controller very quickly highly coordinated behavioral modes emerge, such as crawling or side-rolling, see \fig{fig:snake:behaviors}.
An external perturbation can bring the robot into another behavior.
Without perturbation the robot typically transition into one behavior and remain in it.
We recorded $9\cdot\,10^5$ samples from a typically side-rolling and crawling behavior.

\setlength{\pich}{0.075\linewidth}
\begin{figure}
\begin{center}
  \includegraphics[height=\pich]{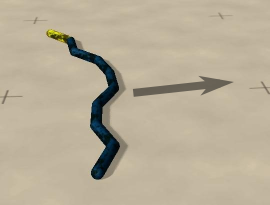}
  \includegraphics[height=\pich]{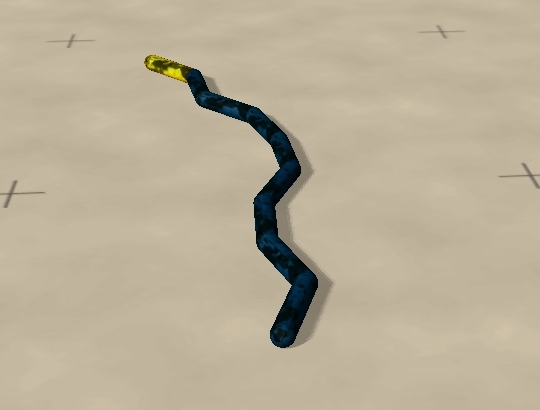}
  \includegraphics[height=\pich]{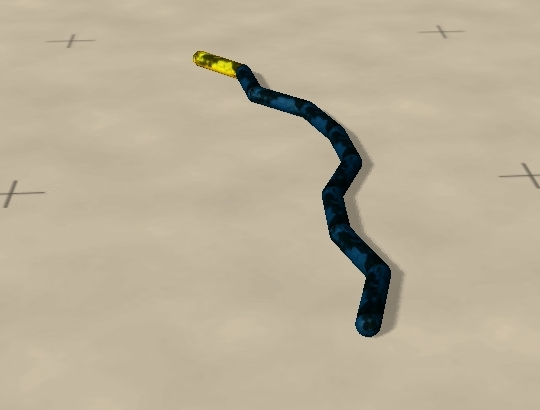}
  \includegraphics[height=\pich]{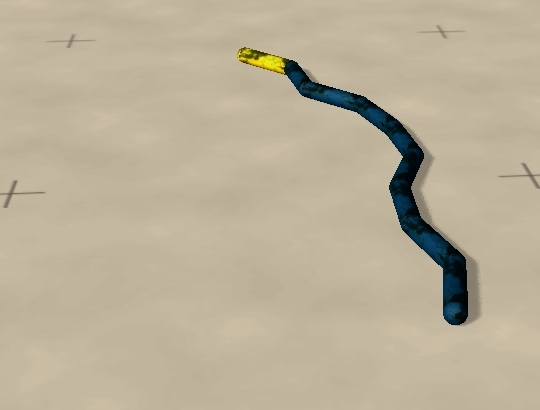}
  \rule[0mm]{.5px}{\pich}
  \includegraphics[height=\pich]{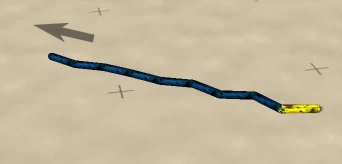}
  \includegraphics[height=\pich]{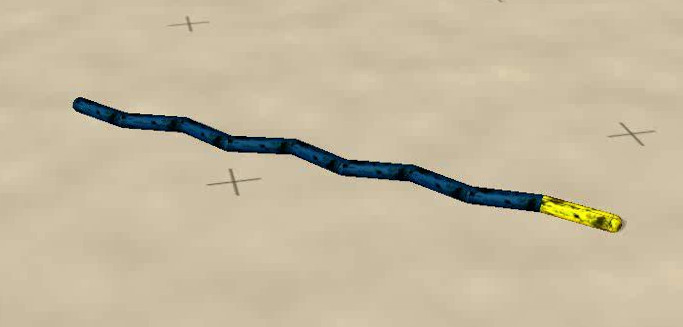}
  \includegraphics[height=\pich]{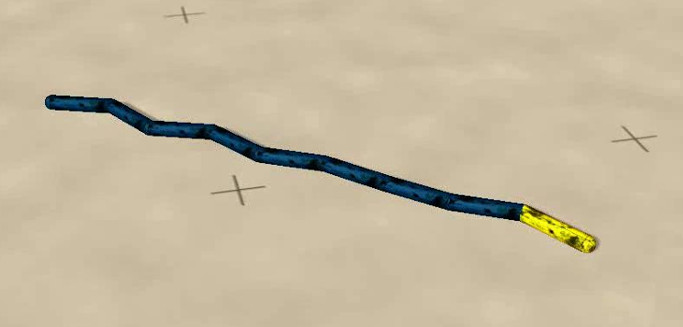}
  \caption{Side-rolling and crawling of the \Snake{}.
    Note that also for rolling each actuator has to act accordingly because no torsion is possible, see \fig{fig:robots}(b).
  }
\label{fig:snake:behaviors}
\end{center}
\end{figure}

\begin{figure}
\begin{center}
  \includegraphics[width=0.154\linewidth]{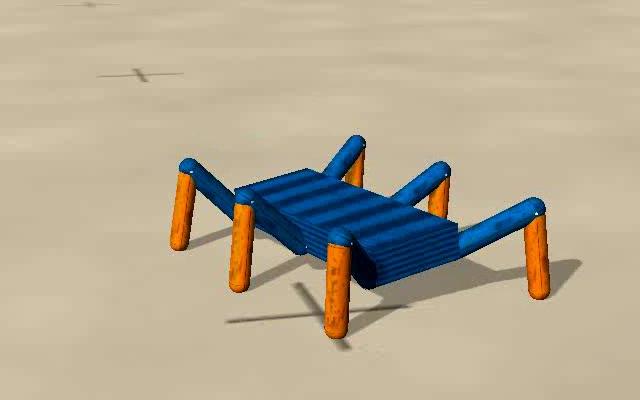}
  \includegraphics[width=0.154\linewidth]{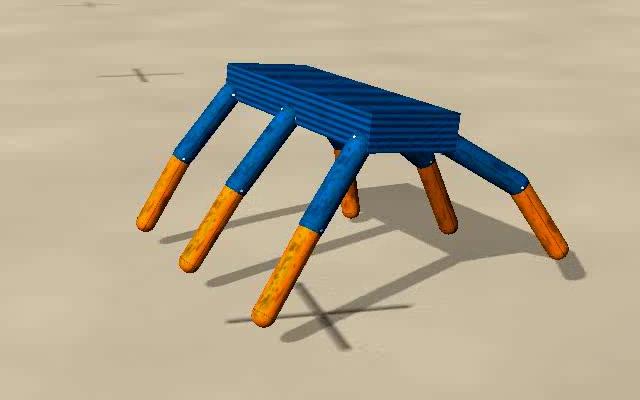}
  \includegraphics[width=0.154\linewidth]{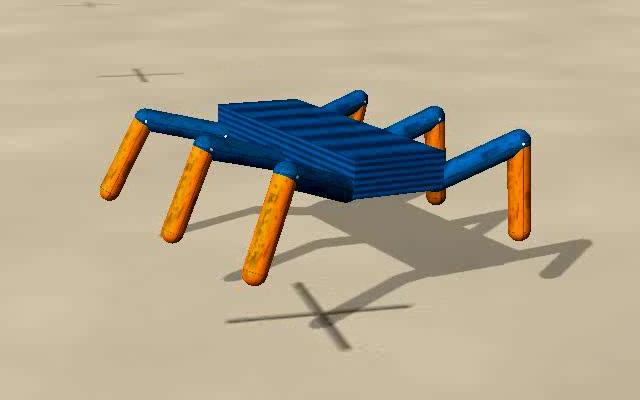}
  \includegraphics[width=0.154\linewidth]{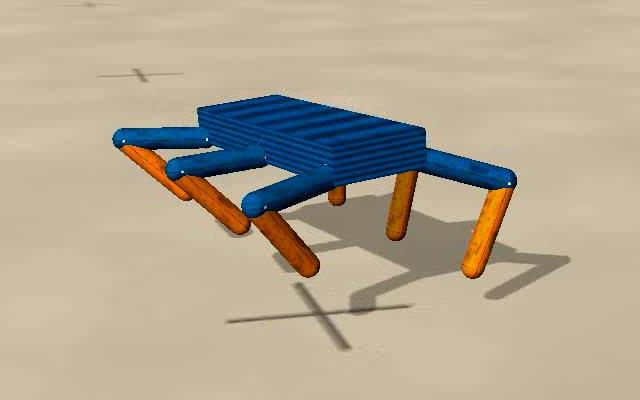}
  \includegraphics[width=0.154\linewidth]{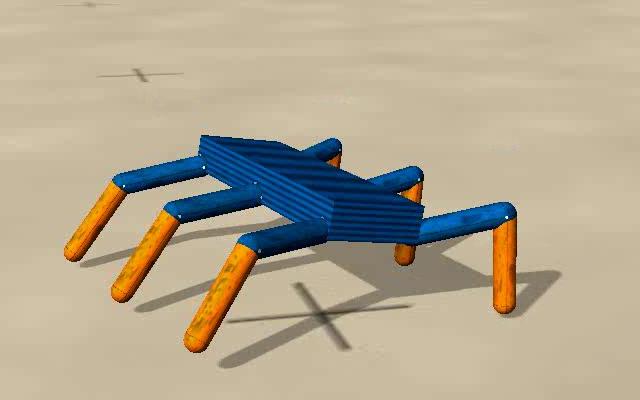}
  \includegraphics[width=0.154\linewidth]{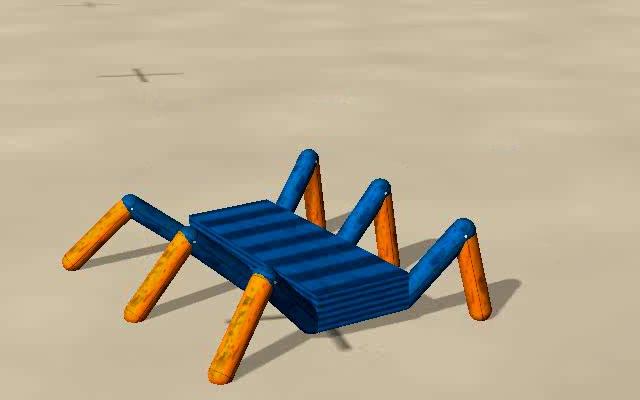}
  \caption{ Jumping motion pattern emerging with vertical speed sensor $8$\, min after
    the start.}
\label{fig:hexapod:behaviors}
\end{center}
\end{figure}

The \Hexapod{} is a insect-like robot with 18\,DoF, see \fig{fig:robots}(a).
The sensor values are the joint angles and the vertical velocity of the trunk.
Note, that the actual phase space dimension is $18+18+4=40$ as we have joint angles, joint velocities, and height and orientation of the trunk.
Controlled with the ULR controller the robot develops a jumping
 behavior, see \fig{fig:hexapod:behaviors}, within 8\,min of simulated time~\cite{DerMartius13}.
The behavior of the robot changes due to the update dynamics of the controller.
For the analysis a stationary process is required, such that we stop the update at some time
and obtain a static controller (fixed $C,h$). Using this the robots behavior stays
 essentially the same.
We analyze three behaviors, where the learning was stopped at 2\,min, 4\,min and 8\,min after the start
 and recorded each $10^6$ samples.
After ``2\,min'' the robot moves periodically up and down.
The ``4\,min'' behavior shows already signs of jumping and the final behavior after ``8\,min''
 is a whole body periodic jumping as displayed in \fig{fig:hexapod:behaviors}.

\subsection{Quantifying the Behavior}\label{s:quant-behavior}

We start with analyzing the data from the \Snake{} experiments followed by the \Hexapod{} experiments.

\subsubsection{\Snake}

\resetsubplot
\begin{figure*}
\begin{center}
  \begin{tabular}{c@{\hspace{-10pt}}c@{\hspace{-.03\columnwidth}}l}
    \subplot{s:s:t} side rolling&\subplot{s:c:t} crawling\\
    \includegraphics[width=0.33\linewidth]{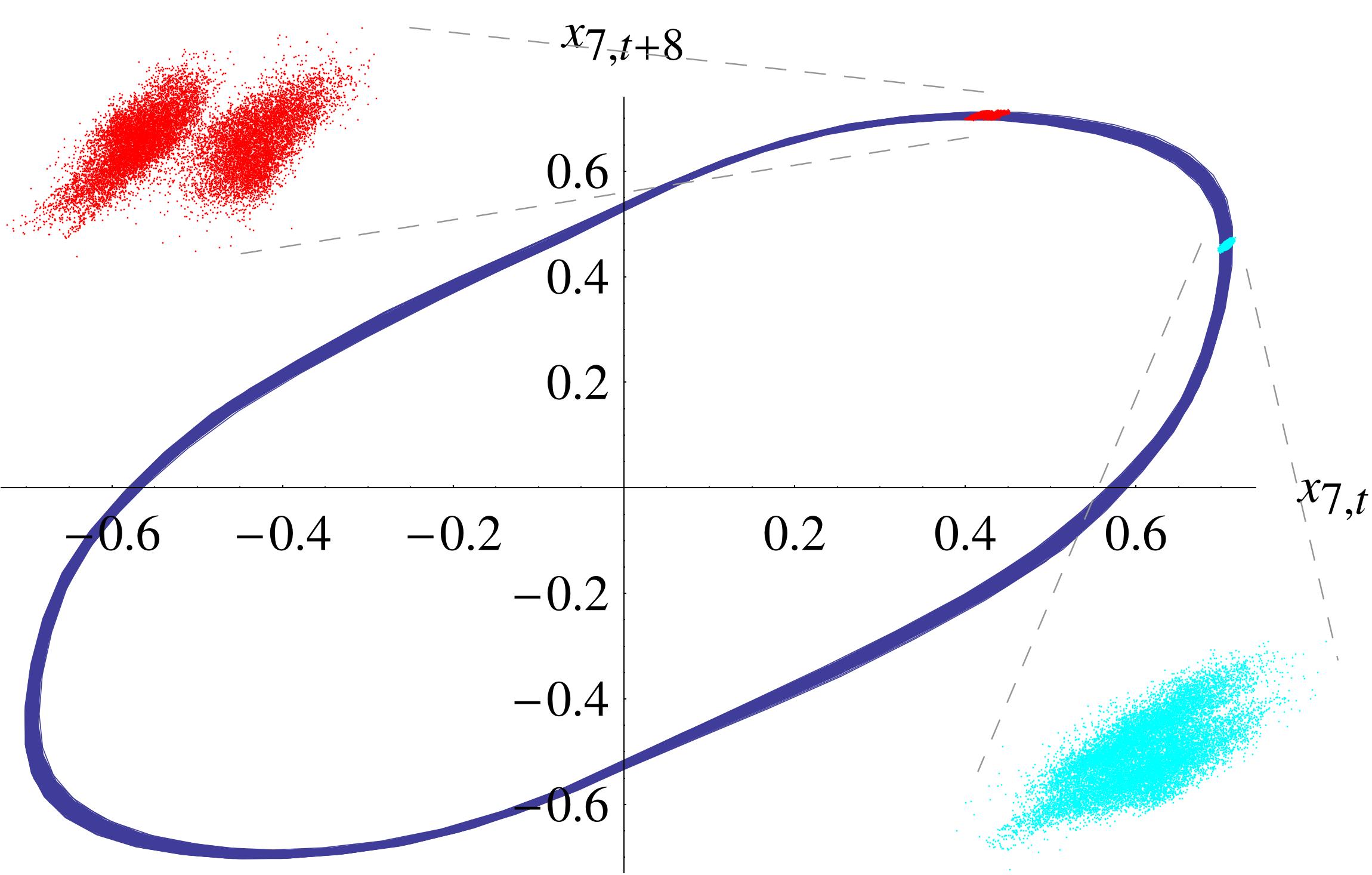}&
    \includegraphics[width=0.33\linewidth]{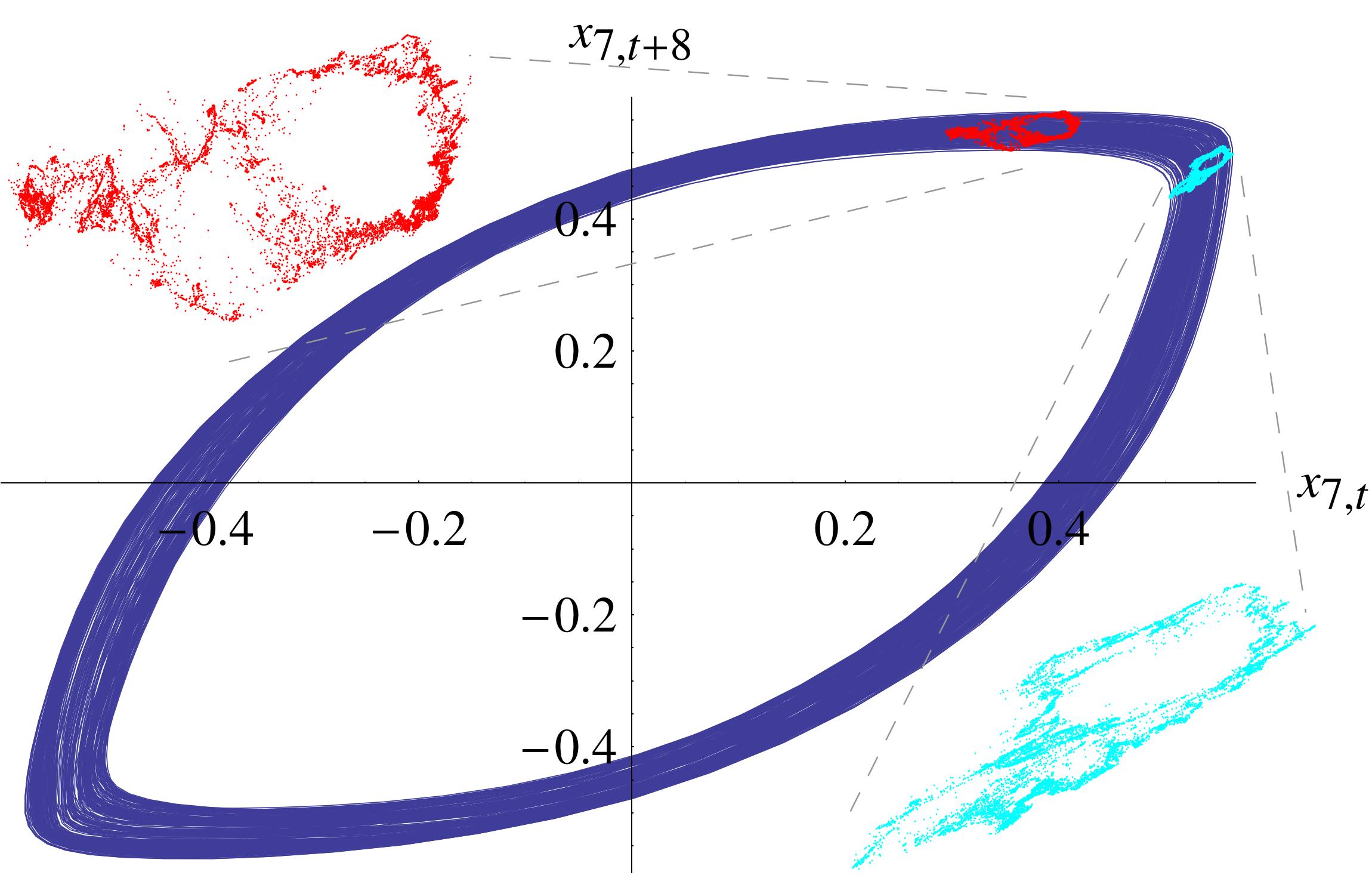}&
          \multirow{3}{*}[1em]{\includegraphics[scale=.9]{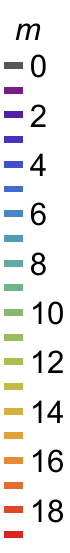}}\\
    \subplot{s:s:d} & \subplot{s:c:d}\\[-.8em]
    \includegraphics[width=0.33\linewidth]{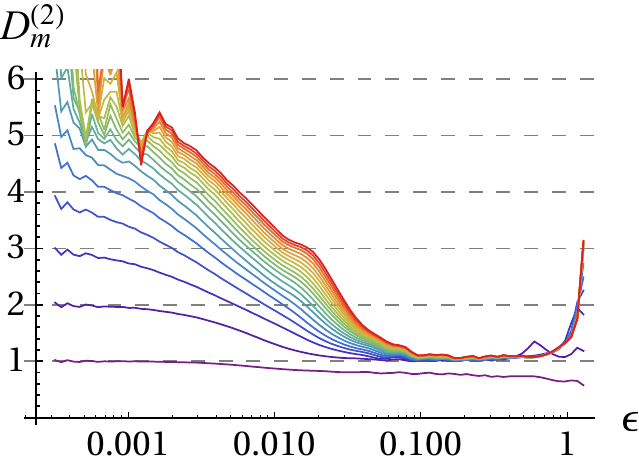}&
    \includegraphics[width=0.33\linewidth]{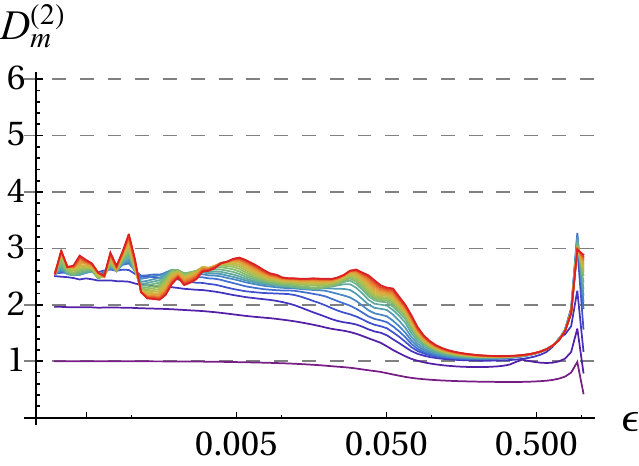}\\[.1em]
    \subplot{s:s:de}& \subplot{s:c:de}\\
    \includegraphics[width=0.43\linewidth]{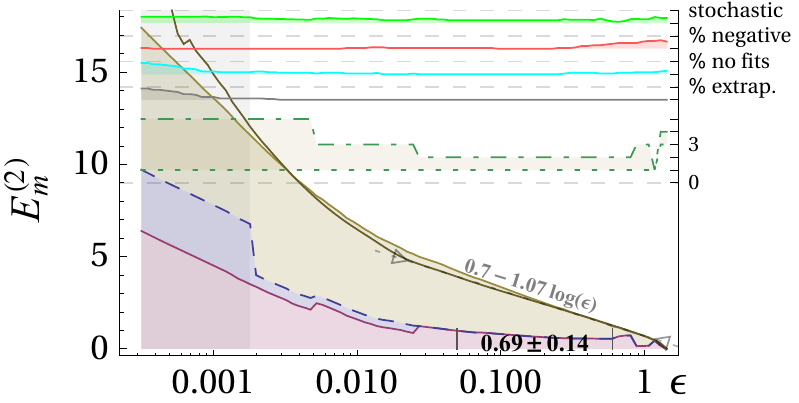}&
    \includegraphics[width=0.43\linewidth]{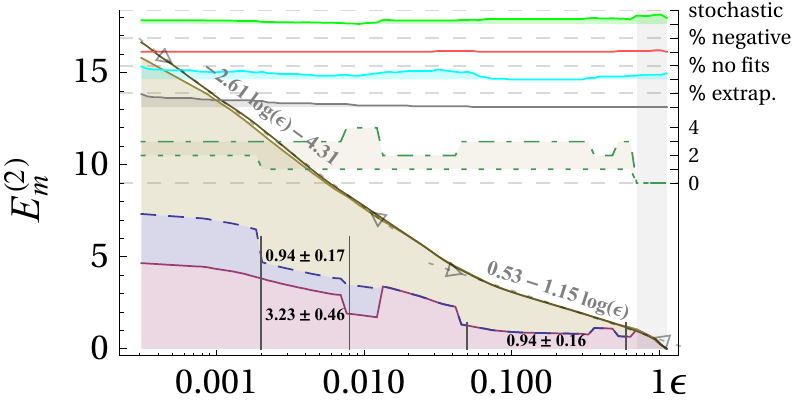}\\
    \multicolumn{3}{c}{\includegraphics[width=.8\columnwidth]{excess-entropy-decomp-legend_with_raw}}\\
    \subplot{s:s:p}& \subplot{s:c:p}\\[-.8em]
    \includegraphics[width=0.33\linewidth]{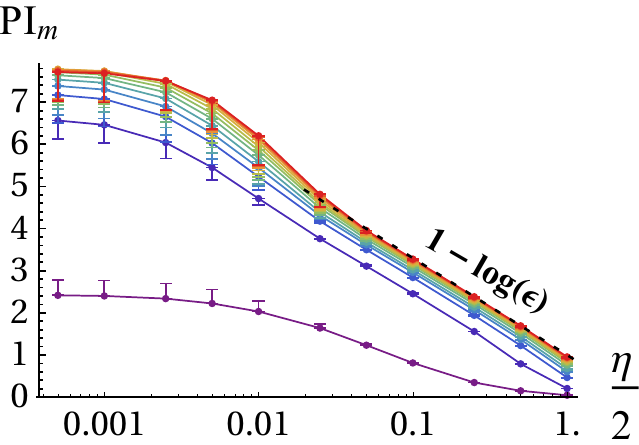}&
    \includegraphics[width=0.33\linewidth]{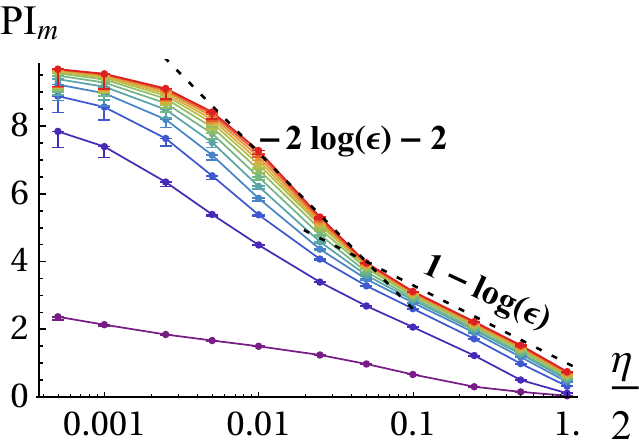}&
      \includegraphics[scale=.9]{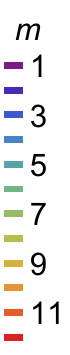}\\
  \end{tabular}
  \caption{Quantification of two \Snake{} behaviors.
    First column: side rolling; second column: crawling (see \fig{fig:snake:behaviors}).
    \infig{\ref{s:s:t},\ref{s:c:t}} Phase plots with Poincar\'e sections at the maximum of first and second embedding.
    \infig{\ref{s:s:de},\ref{s:c:de}} Excess entropy with decomposition (in nats), see \fig{fig:lorenz:decomp} for more details.
    \infig{\ref{s:s:p},\ref{s:c:p}} Predictive information (in nats), see \fig{fig:lorenz:quant} for more details.
    Parameters: delay embedding of sensor $x_7$ (middle segment) with $\tau=8$, $s_{\textrm{min}}=0.05$, $\kappa=0.5$.
 }
\label{fig:snake:quant}
\end{center}
\end{figure*}
In \fig{fig:snake:quant} the result of the quantification of the two behaviors (\fig{fig:snake:behaviors}) are presented. We used the delay embedding of the sensor $x_7$ which is one of the  joint angular sensors of the middle joint.
We find that the side-rolling behavior is roughly 1-dimensional
 on the behaviorally relevant scales ($0.05<\eps<0.5$), see \fig{fig:snake:quant}\ref{s:s:d}.
On smaller scales the dimension raises with each embedding although the embedding dimension is not reached on the observable length scales and therefore the excess entropy does not saturate.
On the larger scales we find the expected scaling behavior \eqnp{eqn:Em-det} with $D\approx 1$.

The crawling behavior is also 1-dimensional on the coarse length scales ($0.1<\eps<0.5$), see \fig{fig:snake:quant}\ref{s:c:d}.
The smaller scales, however, bear a surprise because the dimension raises to a value of $\approx 2.5$.
The excess entropy shows also a slope of $D\approx 1.0$ on the coarse length scales
 and a slope of $D\approx 2.5$ on the smaller scales, see \fig{fig:snake:quant}\ref{s:c:de}.

Calculating the predictive information using the KSG method yields a similar
 result on the large scales, as shown in \fig{fig:snake:quant}\ref{s:s:p},\ref{s:c:p}.
For smaller scales, however, the predictive information always saturates,
 as seen before in \fig{fig:lorenz:quant}\ref{l:0:pi}.
In the crawling case we get again an underestimated dimension of $D\approx 2$.

\begin{table}
  \centering
  \caption{Decomposition of the excess entropy for the \Snake.
    Dimension and unrefined constant, state-, memory-, and core complexity  (in nats) on the coarse deterministic scaling range $\eps \in (0.05,0.6)$.
  }\label{tab:snake:decomp}
  \begin{tabular}{ll|cc}
                      &            & side rolling  & crawling      \\
    \hline
    \eqn{eqn:Em-det}  & $D$        & 1.07          & 1.15          \\
    \eqn{eqn:Em-det}  & $\const$   & 0.7           & 0.53          \\
    \eqn{eqn:EEState} & \EEState   & 0.0$\pm$0 & 0$\pm$0       \\
    \eqn{eqn:EEMem}   & \EEMem     & 0.69$\pm$0.14 & 0.94$\pm$0.16 \\
    \eqn{eqn:EECore}  & \EECore    & 0.69$\pm$0.14  & 0.94$\pm$0.16 \\
  \end{tabular}
\end{table}
The decomposition of the excess entropy follows for the  above experiments are given in \tab{tab:snake:decomp}.
Interestingly, the $\EECore$ attributes a significantly higher complexity to the crawling
 behavior, whereas the plain excess entropy constant would suggest that side-rolling is more complex.
Since we take out the effect of different scales our new quantities are more trustworthy.

Note, that for the crawling behavior at the small scaling range ($\eps <0.01$)
we can also perform the decomposition and obtain $\EEState=0.84\pm 0.17$, $\EEMem=3.23\pm 0.46$,
 however these values are mostly useful for comparison, which we only have on the large scale.
Note that \EEState and \EEMem are not really constant in the plots because the deterministic scaling range is not very good in this case (cf. the quality measure ``no fits'' in Fig.~\ref{fig:snake:quant}\ref{s:c:de}).

\subsubsection{\Hexapod}

\resetsubplot
\begin{figure*}
  \begin{tabular}{c@{\hspace{-0.01\columnwidth}}c@{\hspace{-0.01\columnwidth}}c@{\hspace{-0.01\columnwidth}}l}
    \subplot{h:2:t} 2\,min&\subplot{h:4:t} 4\,min&\subplot{h:8:t} 8\,min \\
    \includegraphics[width=0.3\linewidth]{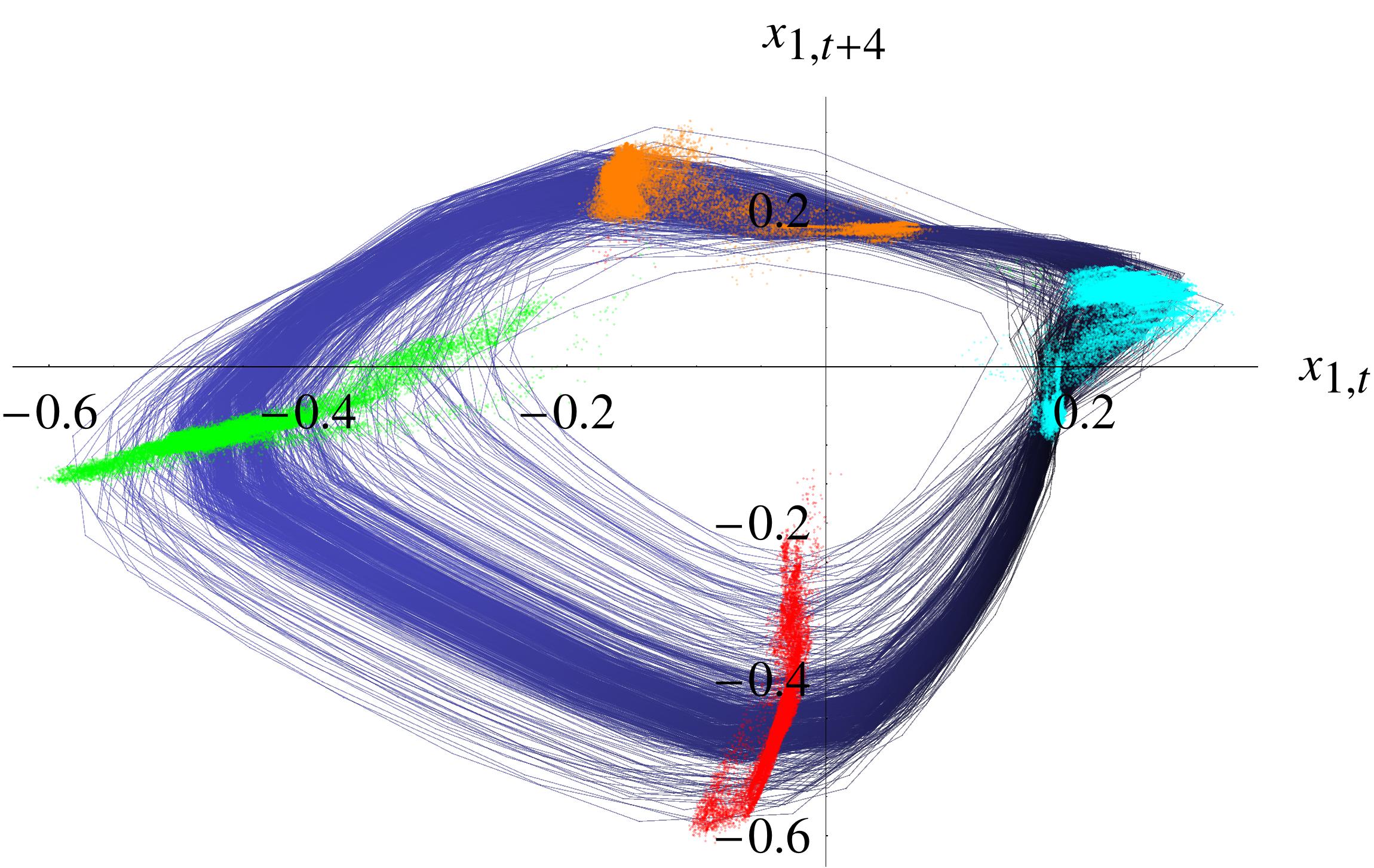}&
    \includegraphics[width=0.3\linewidth]{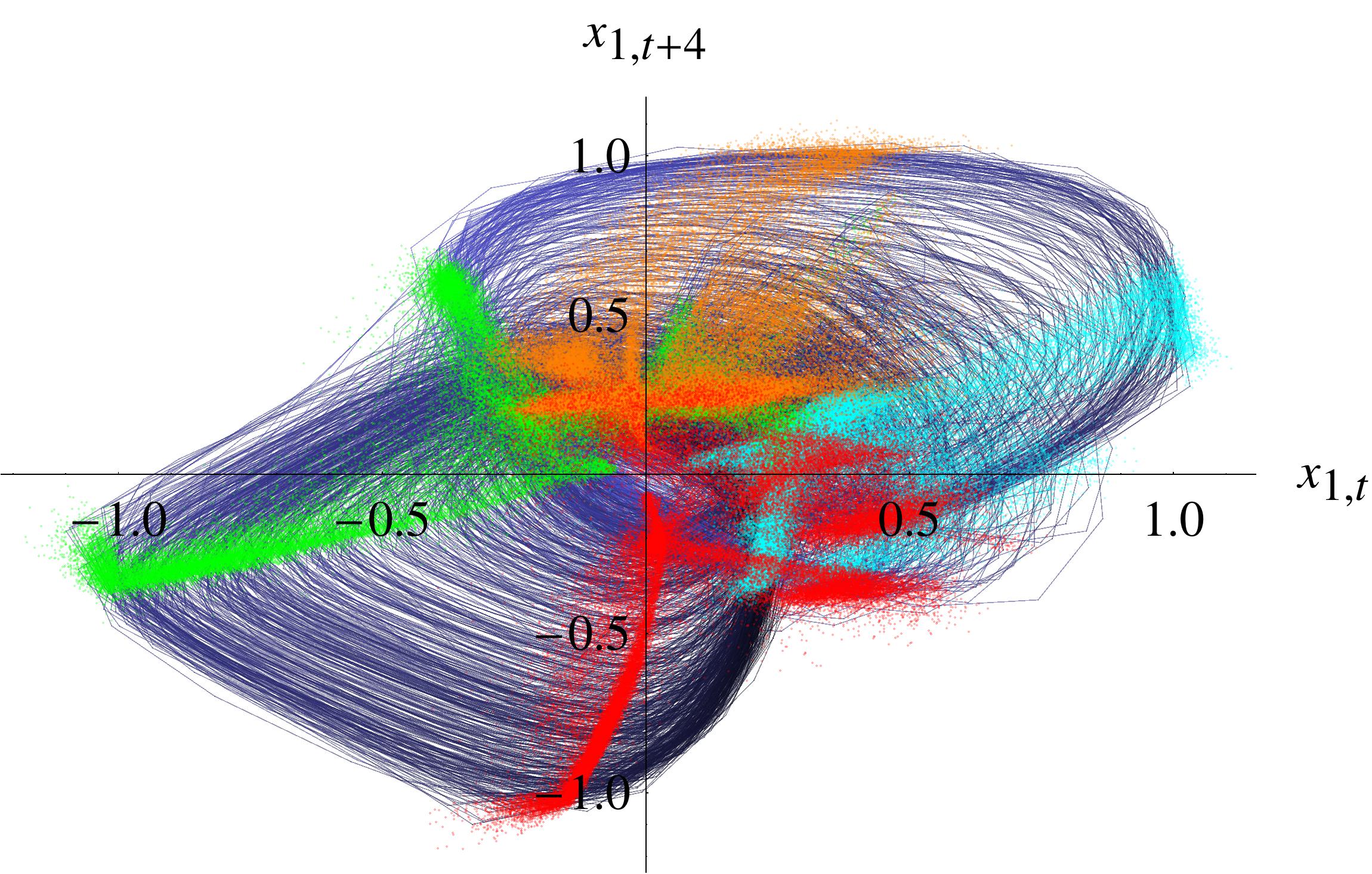}&
    \includegraphics[width=0.3\linewidth]{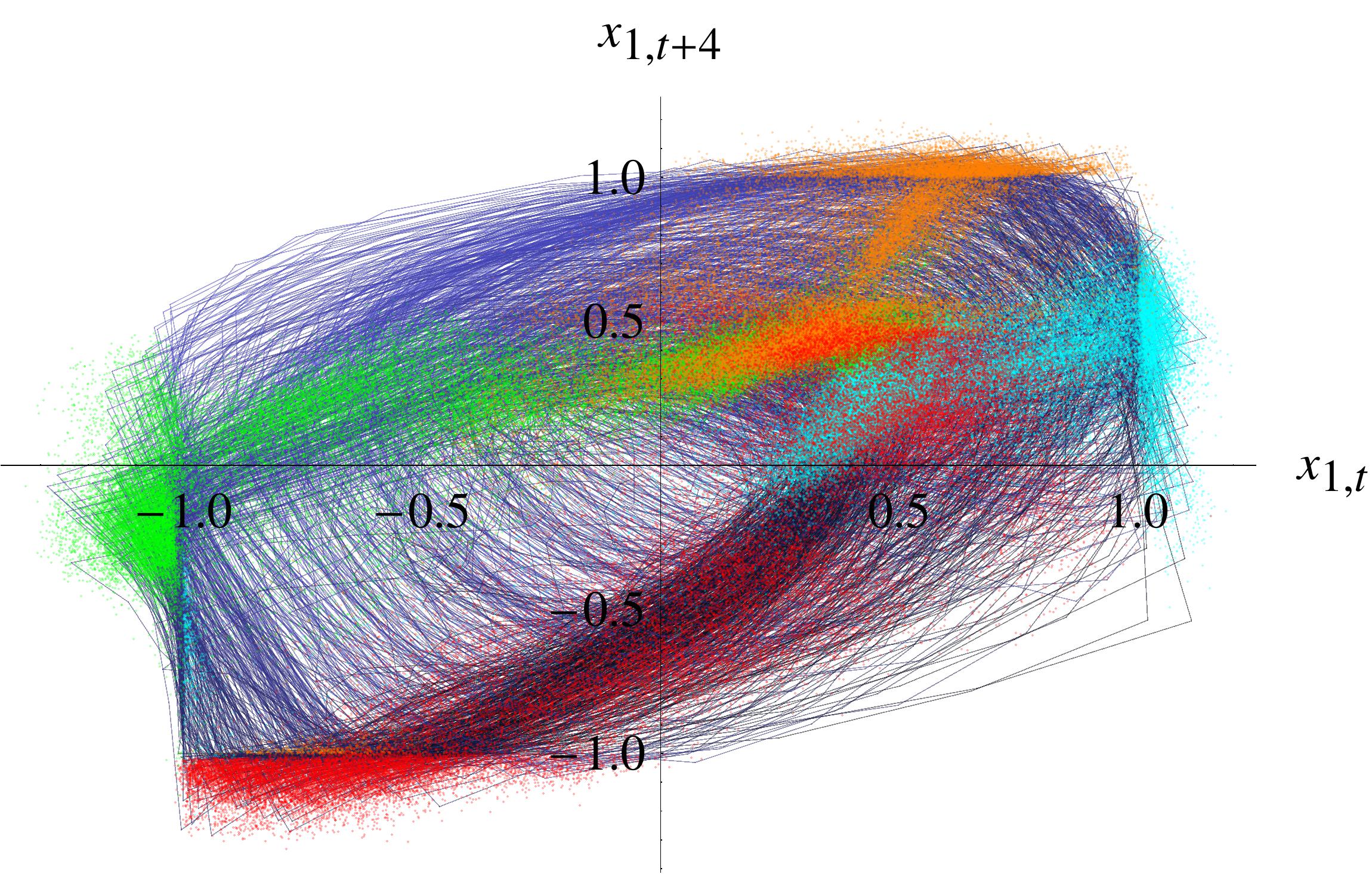}&\multirow{3}{*}[1em]{\includegraphics[scale=.75]{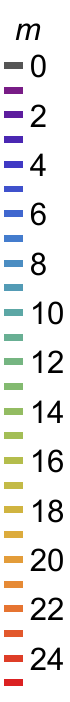}}\\
    \subplot{h:2:d} &\subplot{h:4:d} &\subplot{h:8:d}  \\[-.8em]
    \includegraphics[width=0.3\linewidth]{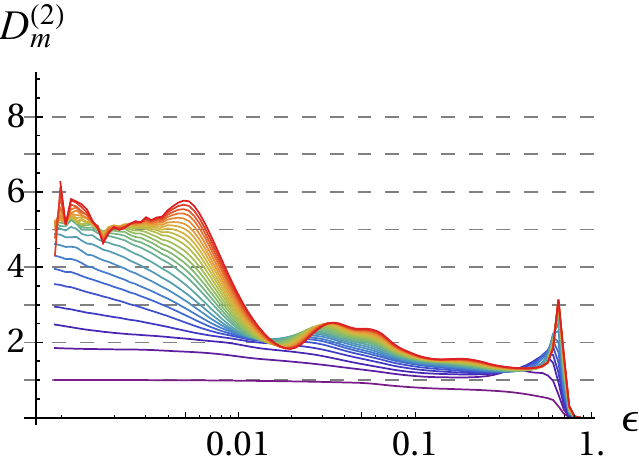}&
    \includegraphics[width=0.3\linewidth]{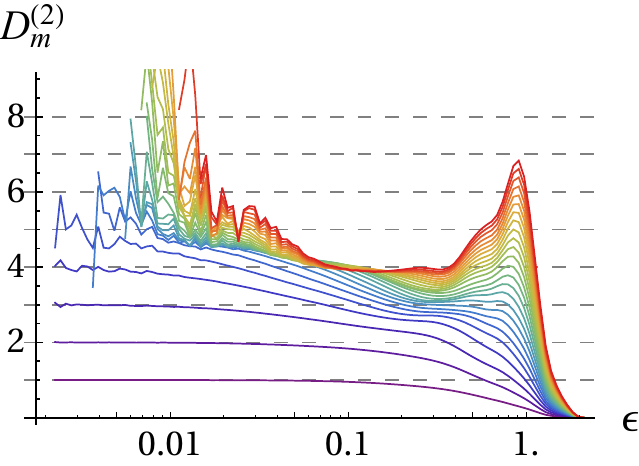}&
    \includegraphics[width=0.3\linewidth]{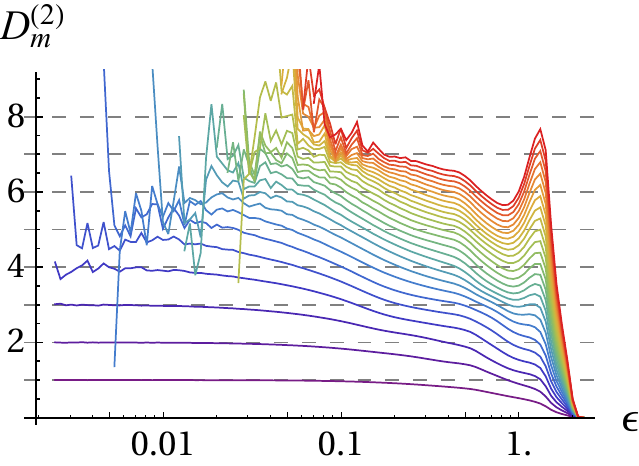}\\
    \subplot{h:2:de} &\subplot{h:4:de} &\subplot{h:8:de}  \\
    \includegraphics[width=0.34\linewidth]{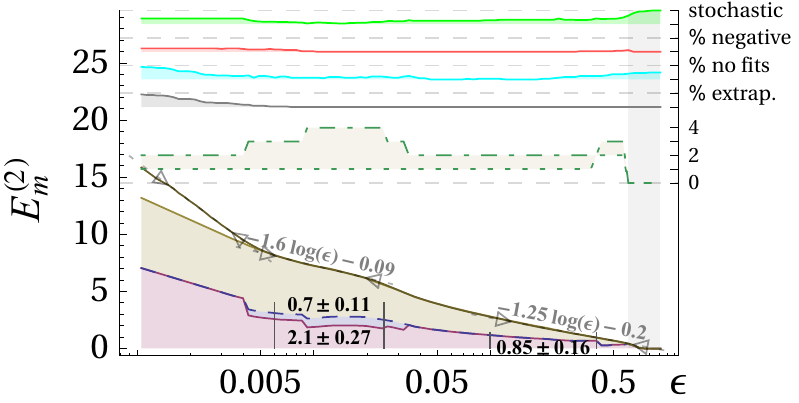}&
    \includegraphics[width=0.34\linewidth]{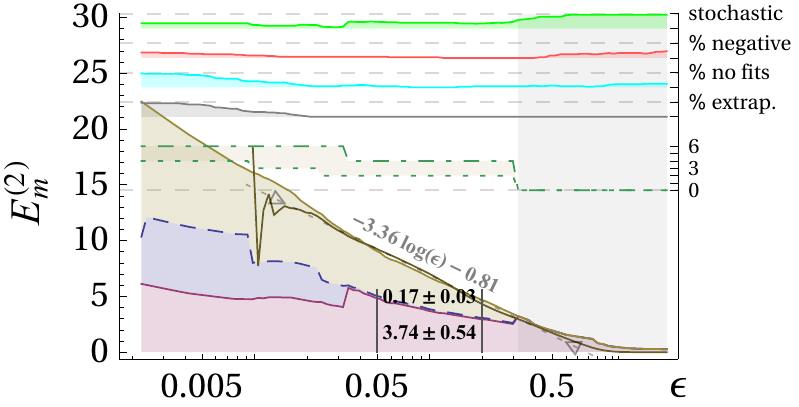}&
    \includegraphics[width=0.34\linewidth]{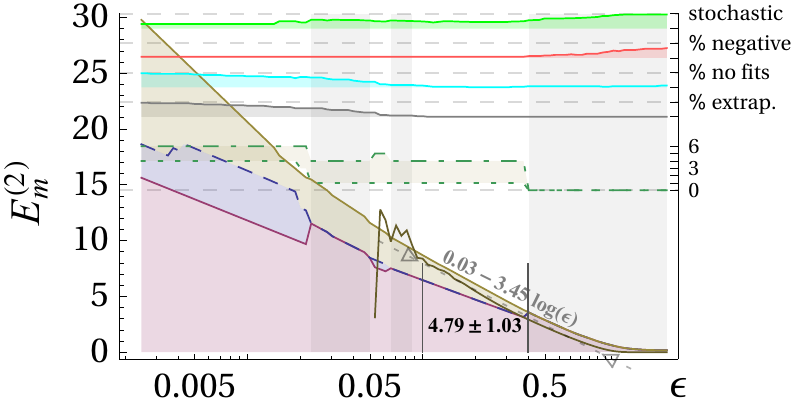}\\
    \multicolumn{3}{c}{\includegraphics[width=.9\columnwidth]{excess-entropy-decomp-legend_with_raw}}\\
    \subplot{h:2:pi} &\subplot{h:4:pi} &\subplot{h:8:pi}&\multirow{2}{*}{\includegraphics[scale=.75]{legend12-1}} \\[-.8em]
    \includegraphics[width=0.3\linewidth]{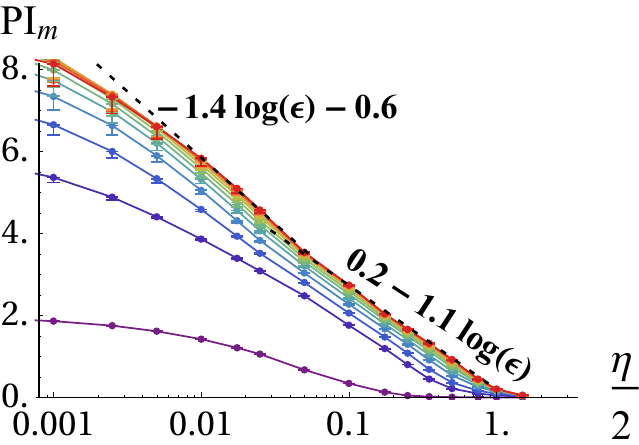}&
    \includegraphics[width=0.3\linewidth]{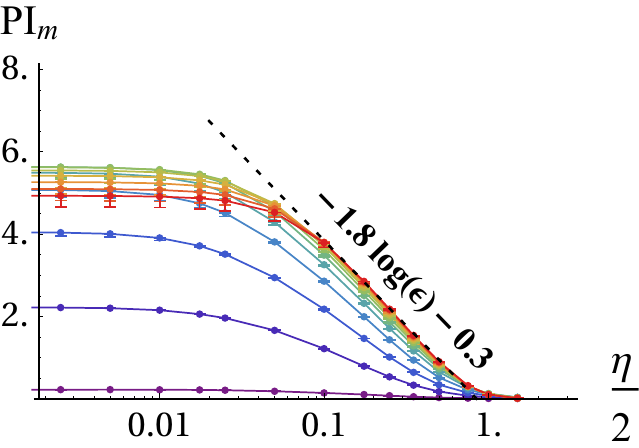}&
    \includegraphics[width=0.3\linewidth]{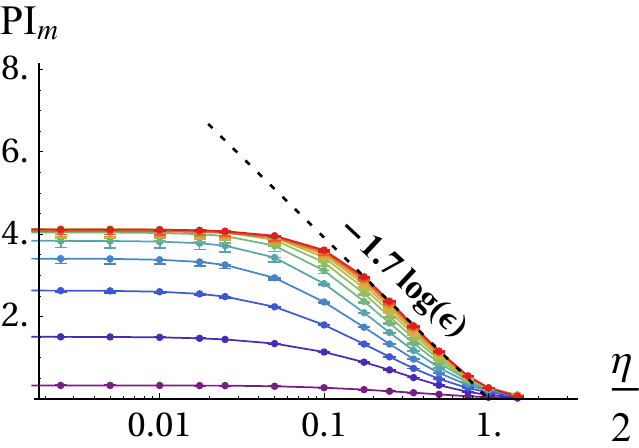}\\
  \end{tabular}
  \caption{Quantification of the three \Hexapod{} behaviors.
    Behaviors: first column: 2\,min; second column: 4\,min; third column 8\,min, see \fig{fig:hexapod:behaviors}.
    \infig{\ref{h:2:t}--\ref{h:8:t}} Phase plots with Poincar\'e sections at the maximum and minimum of first and second embedding. The color of the trajectory (blue--black) encodes for the third embedding dimension.
    \infig{\ref{h:2:de}--\ref{h:8:de}} Excess entropy with decomposition (in nats), see \fig{fig:lorenz:decomp} for more details.
    The excess entropy fit at the smallest scale in \ref{h:2:de} is $-18-5\log(\eps)$.
    \infig{\ref{h:2:pi}--\ref{h:8:pi}} Predictive information (in nats), see \fig{fig:lorenz:quant} for more details.
    Parameters: delay embedding of sensor $x_1$ (up-down direction of right hind leg) with $\tau=4$, $s_{\textrm{min}}=0.05$, $\kappa_{\textrm{max}}=0.5$.
  }
\label{fig:hexapod:quant}
\end{figure*}
\Fig{fig:hexapod:quant} shows the quantification of the three successive behaviors of the \Hexapod{} robot,
 reconstructed from the delay embedding of sensor $x_1$ (up-down direction of right hind leg).
We clearly see an increase in dimensionality of the behavior, \fig{fig:hexapod:quant}\ref{h:2:d}--\ref{h:8:d}
 which we can expect because the high jumping behavior requires more compensation movements due to different landing poses.
For the first behavior we can identify three scaling ranges of different dimensionality and complexity.
On the coarse level $[0.1,0.6]$ the behavior is described by a dimension of $1.2$
whereas on medium scales $[0.006,0.02]$
 we find an approximately $1.6$ dimensional behavior and on the small scales $[0.001,0.005]$ it appears 5 dimensional.

For the second behavior we find only a short plateau in \fig{fig:hexapod:quant}\ref{h:4:d} at $[0.06,0.3]$ of dimensionality 4.
The excess entropy shows a lower dimensionality of $\approx 3.4$, see \fig{fig:hexapod:quant}\ref{h:4:de}.
For the third behavior we cannot identify a clear plateau in the dimension plot.
The excess entropy allows us to attribute a dimensionality and core complexity
 to the behavior on large scales, as summarized in \tab{tab:hexapod:constants}.
Whereas the scaling behavior of the excess entropy suggests a similar dimensionality, the
 core complexity suggest that the 8\,min behavior is more complex.

The predictive information calculated with the KSG method underestimates the dimension
 by far and is only able to report structure on the coarsest scale, see \fig{fig:hexapod:quant}\ref{h:2:pi}--\ref{h:8:pi} and compare for instance slope $1.8$ of PI in \ref{h:4:pi} with slope $3.36$ of $\Etwo$ in \ref{h:4:de}.
Note that usually only the value of the mutual information without added noise (or only very small one)
is reported (left-most value in our graphs).
In our case this would attribute the highest complexity to the 2\,min behavior and the lowest to the 8\,min behavior, see also \tab{tab:hexapod:constants}. This is because the intrinsic noise level increases from 2\,min to 8\,min, such that
 the plateau level is much lower. 

\begin{table}
  \centering
  \begin{tabular}{ll|cccc}
                       &            & 2\,min (fine)  & 2\,min (coarse)& 4\,min (coarse) & 8\,min(coarse)\\
    \hline
                      &\eps-range   &$(0.006, 0.025)$&$(0.1, 0.4)$    &$(0.05, 0.2)$   & $(0.1, 0.4)$\\
    \eqn{eqn:Em-det}&$D$            & 1.6            & 1.25           &  3.36         & 3.45\\
    \eqn{eqn:Em-det}&$\const$       & -0.09          & -0.2           &  -0.81        & 0.03\\
    \eqn{eqn:EEState}& \EEState     &  0.7$\pm$0.11  &  0$\pm$0       & 0.17$\pm$0.03 & 0$\pm$0\\
    \eqn{eqn:EEMem}  & \EEMem       &  2.1$\pm$0.27  &  0.85$\pm$0.16 & 3.74$\pm$0.54 & 4.79$\pm$1.03\\
    \eqn{eqn:EECore} & \EECore      &  2.8$\pm$0.38  &  0.85$\pm$0.16 & 3.91$\pm$0.57 & 4.79$\pm$1.03\\
    \hline
    \eqn{eqn:PI}     & $\PI (\eta=0)$ & $ \approx 9$   &               & $\approx 5.5$  & $\approx 4$\\
  \end{tabular}
  \caption{Decomposition of the excess entropy for the \Hexapod{} behaviors on the fine and coarse scale.
    For the 4\,min and 8\,min behavior we have no reliable estimate on the fine scale.
    The core complexity and the dimension rises from 2\,min to 4\,min.
    From 4\,min to 8\,min the dimension stayed the same but the core complexity increases.
    The last line provides the predictive information on the smallest scale.
   }\label{tab:hexapod:constants}
\end{table}

\section{Discussion} \label{s:Discussion}

In recent years research in autonomous robots has been more and more successful in developing algorithms for generating behavior from generic task-independent objective functions~\cite{ay08:predinf_explore_behavior,ZahediAyDer2010:HigherCoordination,MartiusDerAy2013,Klyubin2005:Empowerment,SalgeGlackinPolani2014:EmpowermentIntro,schmidhuber:02predictable,oudeyer:05c,FrankSchmidhuber2013:curiosity}.
This is valuable, on the one hand, as a step towards understanding and creating autonomy in artificial beings,
 and on the other hand, to obtain eventually more effective learning algorithms
 for mastering real-life problems.
This is because the task-agnostic drives aim at efficient exploration of the agent's
 possibilities by generically useful
 behavior~\citep{Schmidhuber91:CuriousControl,oudeyer:05c,oudeyer07:IntrinsicMotivation,DerMartius11,MartiusDerAy2013}
 and at generating an operational state of the system~\cite{DoncieuxMouret2014:review-ER}
 from which specific tasks can quickly be accomplished.

Very often emerging behaviors are not the result of optimizing a global function but
for instance self-organize from local interaction rules.
We propose to use information theoretic quantities for quantifying these behaviors.
There are already studies using information theoretic measures to characterize emergent behavior of autonomous robots~\cite{lungarella06:InfoFlowInSML,SchmidtHoffmann2013:BoostrappingPerception,WangLizieretal:quantifyswarms} --- in particular with the aim to use them as task-independent objective functions~\cite{ay08:predinf_explore_behavior,ZahediAyDer2010:HigherCoordination,MartiusDerAy2013,Klyubin2005:Empowerment,SalgeGlackinPolani2014:EmpowermentIntro,schmidhuber:02predictable}.
Usually these information theoretic quantities are estimated for a fixed partition, \ie a specific discretization, or by using differential entropies assuming that the data were sampled from a probability density. In this paper we show that it is both necessary and fruitful to explicitly take into account the scale dependence of the information measures.

We quantify and compare behaviors emerging from self-organizing control of high-dimensional robots by estimating the excess entropy and the predictive information of their sensor time series.
The estimation is done on the one hand using entropy estimates based on the correlation integral and
on the other hand the predictive information is directly estimated as a mutual information using an estimator (KSG) based on differential entropies.
The latter implicitly assumes stochastic data, \ie the existence of a smooth probability density for the temporal joint distribution,
such that the predictive information converges towards a finite value.
However, this is not the case for deterministic systems.
Due to the adaptivity of the algorithm the predictive information is then calculated at a certain length scale which depends on the available amount of data.
Thus, a naive application of the KSG estimator can lead to totally meaningless results, because it does not allow to control for the length scale. To avoid this problem and to control the length scale, we applied additive noise.
Nevertheless, our results show that the complexity estimates based on the KSG estimator are much less reliable
 than those based on the correlation integral.
While for simple stochastic systems such as autoregressive models both estimators deliver consistent results for small embedding dimensions,
 the predictive information is underestimated due to finite sample effects for larger embedding dimensions (cf.~\fig{fig:lorenz:armodel}).
For more complex systems such as the Lorenz attractor or the behavioral data from robots the KSG algorithm  underestimates the dimensionality, see \eg \fig{fig:hexapod:quant}\ref{h:2:pi}--\ref{h:8:pi},
 does not allow to extract length scales properly, and blurs the distinction between stochastic and deterministic scaling ranges.

An important result of the presented study is that the complexity of the behavior can be different at different length scales: an example is the crawling behavior of the \Snake which is 1-dimensional on the coarse scale and 2.6-dimensional on the small scales, see \fig{fig:snake:quant} (the same for the \Hexapod).
In order to quantify these effects we analyzed for the first time the resolution dependence of the well-known complexity measure for time series known as excess entropy \cite{shaw84,Crutchfield03}, effective measure complexity \cite{Grassberger86}, or predictive information \cite{bialek01}.
We show that the scale dependent excess entropy reflects different properties of the system: scale, dimension and correlations. For deterministic systems the attractor dimension controls how the complexity increases with increasing resolution,
 while the offset depends on the overall scale of the behavior, \eg the amplitude of an oscillation, and on the structure of the temporal correlations. In order to disentangle the latter two effects we introduced a new decomposition of the excess entropy with the \emph{core complexity} quantifying the structural complexity beyond the dimension. In order to estimate it one has to remove the effect of the overall scale of the behavior from the excess entropy.
We devised an algorithm based on an automatic identification of scaling ranges in the conditional mutual information $\delta h_m(\epsilon)$.

 For noisy systems the excess entropy remains finite.
 The actual value, however, depends on the noise level, which is consistent with the intuition: The larger the noise the lower the complexity if all other factors remain constant.
While we have already a good understanding of these quantities for systems with clearly identifiable scaling regions as in the noisy Lorenz system, very often real systems do not show such ideal types of behavior.
Therefore we had to use heuristic procedures which provide plausible results for the data under study.
Our quantification of the behaviors of the \Snake found
 that crawling has a higher complexity than side-winding,
 even though the original excess entropy would have suggested the opposite. 
For the \Hexapod as a learning system (aiming at maximizing time local PI) we were able to quantify the learning progress: The decomposition of the excess entropy revealed the strategy used to increase the complexity:
(1) The dimensionality of the behavior increased on the large scales.
This does not necessarily mean that the attractor dimension in the mathematical sense increased, because the latter is related to behavior on infinitely small scales which might be irrelevant for the character of the behavior.
(2) The core complexity is increased on the large scales indicating also more complex temporal correlations in the learned behavior.
We also discuss further strategies for increasing the complexity of a time-series, that could be employed in designing future
 learning algorithms, most notably, the decrease in the entropy rate and the increase of the state complexity.

Our study contributes a well grounded and practically applicable quantification measure for the behavior of autonomous systems
 as well as time-series in general.
In addition, a set of quality indicators are provided for self-assessment in order to avoid spurious results.

Future work is to develop a better theoretical understanding of the scale dependence of the excess entropy for realistic behaviors that are not low dimensional deterministic or linearly stochastic such as the behavior observed in the \Hexapod after 8 minutes of learning, \fig{fig:hexapod:quant}. We expect also further insights by applying the same kind of analysis to movement data from animals and humans.


\subsection*{Acknowledgments}

This work was supported by the DFG priority program 1527 (Autonomous Learning) and by the European Community's Seventh Framework Programme (FP7/2007-2013) under grant agreement no.~318723 (MatheMACS)
 and from the People Programme (Marie Curie Actions) of the European Union's Seventh Framework Programme (FP7/2007-2013) under REA grant agreement no.~291734.


\bibliographystyle{abbrv}
\setlength{\bibsep}{1pt plus 0.3ex}

\newpage
\appendix
\section{Comparison and validation of estimators using autoregressive model of the Lorenz attractor}\label{sec:AR2}

In this appendix we compare the estimates based on the correlation integral with the estimates using the KSG algorithms for a system for which the excess entropy can be computed analytically: an
autoregressive (AR) model of the Lorenz attractor, which results in a stochastic system with defined correlation structure.
The dimension of a stochastic systems is infinite, \ie for embedding $m$ the correlation integral yields the full embedding dimension.
However, the excess entropy stays finite.

The AR model of a time series $x_i$ is given by
\begin{align}
  x_{n+1} &= \sum^p_{k=0} a_k x_{n-k} + \xi_n \label{eqn:arm}
\end{align}
where $p$ is the order and the $a_k$ are the parameters which are fit to the data and $\xi_n$ independent identically distributed Gaussian noise.

Let us consider the example of an AR2 (p=2) process
\begin{align}
  x_{n+1} &= a_1 x_{n} + a_2 x_{n-1} + \xi_n
\end{align}
and calculation the mutual information of subsequent time steps:
\begin{align}
 I \left(X_{n+1}:X_n\right) & = 2H\left(X_n\right) -H\left(X_{n+1}X_n\right) \\
 H\left(X_n\right) &= \frac 1 2 \ln \left(2 \pi  e \left\langle x_n^2\right\rangle \right) \\
 \left\langle x^2\right\rangle &= a_1^2\left\langle x^2\right\rangle  + a_2^2\left\langle x^2\right\rangle  + \left\langle \xi ^2\right\rangle \\
 \beta  &:= \left\langle x^2\right\rangle = \sigma ^2/\left(1-a_1^2-a_2^2\right) &\text{with } \sigma^2 := \left\langle \xi^2\right\rangle\\
 H\left(X_n\right) &= \frac 1 2 \ln (2 \pi  e \beta ) \\
 H\left(X_{n+1}X_n\right) &= \frac 1 2 \ln \left((2 \pi e)^2 |K|\right)\\
K&= \begin{pmatrix} \beta  & \left\langle x_{n+1}x_n\right\rangle  \\
   \left\langle x_{n+1}x_n\right\rangle  & \beta \end{pmatrix}\\
\left\langle x_{n+1}x_n\right\rangle &= a_1 \left\langle x_n^2\right\rangle  + a_2 \left\langle x_{n-1}x_n\right\rangle\\
     &= \frac{a_1}{\left(1-a_2\right)}\beta  = r \beta &\text{with } r:=a_1/\left(1-a_2\right)\\
H\left(X_{n+1}X_n\right) &= \frac 1 2 \ln \left((2 \pi  e )^2 \beta^2\left(1-r^2\right)\right)\\
I\left(X_{n+1}:X_n\right) &= \frac 1 2 \ln  \left(1-r^2\right)
                           = \frac 1 2 \ln \left(1-\left(\frac{a_1}{1-a_2}\right)^2\right)\,.
 \end{align}

For larger time horizons it suffices to calculate $I \left(X_{n+1}X_n:X_{n-1}X_{n-2}\right)$ because
 of the Markov property more terms will not increase $I$.

\begin{align}
 I\left(X_{n+1}X_n:X_{n-1}X_{n-2}\right)
 &= 2H\left(X_{n+1}X_n\right) -H\left(X_{n+1}X_nX_{n-1}X_{n-2}\right) \\
 H\left(X_{n+1}X_nX_{n-1}X_{n-2}\right) &= \frac{1}{2}\ln \left((2 \pi  e )^4 |K|\right) \\
  K&=\begin{pmatrix}
  \beta  & r_1 \beta  & r_2 \beta  & r_3 \beta  \\
  r_1 \beta  & \beta  & r_1 \beta  & r_2 \beta  \\
  r_2 \beta  & r_1 \beta  & \beta  & r_1 \beta  \\
  r_3 \beta  & r_2 \beta  & r_1 \beta  & \beta
 \end{pmatrix}\label{eqn:AR2MIK}\,.
\end{align}
Where the $r_i$ are the correlation coefficient for $i$-step delay ($r_1 = r$).
\begin{align}
  \left\langle x_{n+1}x_{n-1}\right\rangle  &= a_1 \left\langle x_nx_{n-1}\right\rangle  + a_2\left\langle
    x_{n-1}^2\right\rangle =a_1r \beta  + a_2\beta  = \beta  r_2 \\
  r_2&= a_1r + a_2 \\
 \left\langle x_{n+1}x_{n-2}\right\rangle &= a_1 \left\langle x_nx_{n-2}\right\rangle  + a_2\left\langle
    x_{n-1}x_{n-2}\right\rangle =a_1r_2\beta + a_2r \beta  = \beta  r_3 \\
  r_{3}&=a_1r_2+ a_2r\,.
\end{align}
Thus the 2-step mutual information is given by:
\begin{align}
 I\left(X_{n+1}X_n:X_{n-1}X_{n-2}\right)&=\frac 1 2 \ln \frac{1}{\left(1+a_2\right)^2 \left(\left(
 a_2-1\right)^2 - a_1^2\right)}\,.
\end{align}

If the process is observed with delay $\tau$, \ie $y_n = x_{\tau \cdot n}$
 then only the $r_i$ in \eqn{eqn:AR2MIK} have to be substituted by $r_{\tau \cdot i}$
 because $\beta$ cancels. For convenience the list of the first $r_i$ follows:
\begin{align*}
  r_1 &= a_1/(1 - a_2) \qquad\qquad r_2 = a_1^2/(1 - a_2) + a_2 \\
  r_3 &= (a_1 (-a_1^2 + (-2 + a_2) a_2))/(-1 + a_2)\\
  r_4 &= (-a_1^4 + a_1^2 (-3 + a_2) a_2 + (-1 + a_2) a_2^2)/(-1 + a_2)\\
  r_6 &= (-a_1^6 + a_1^4 (-5 + a_2) a_2 + 3 a_1^2 (-2 + a_2) a_2^2 + (-1 + a_2) a_2^3)/(-1 + a_2).
\end{align*}
For larger delays values the expressions become very large
 such that it becomes intractable to compute the MI for $\tau>10$ (requires $r_{30}$)
 because numerical errors accumulate even after algebraic simplification.
The supplementary (\href{http://playfulmachines.com/QuantBeh2015/}{\tt{\small playfulmachines.com/QuantBeh2015}}) contains a Mathematica file for calculating the MI's up to $\tau=10$.

\begin{figure}
   \centering
  \begin{tabular}{cc@{\hspace{-1pt}}l}
    AR2 -- delay 1 & AR2 -- delay 10\\
    \subplot{a:1:d2} & \subplot{a:10:d2}\\[-1.1em]
    \includegraphics[width=.3\columnwidth]{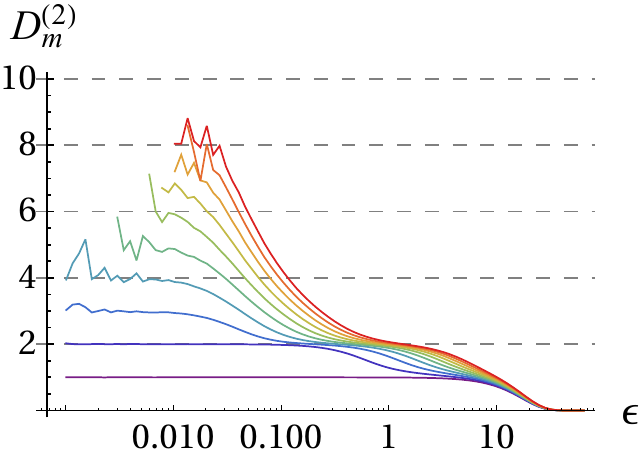}&
    \includegraphics[width=.3\columnwidth]{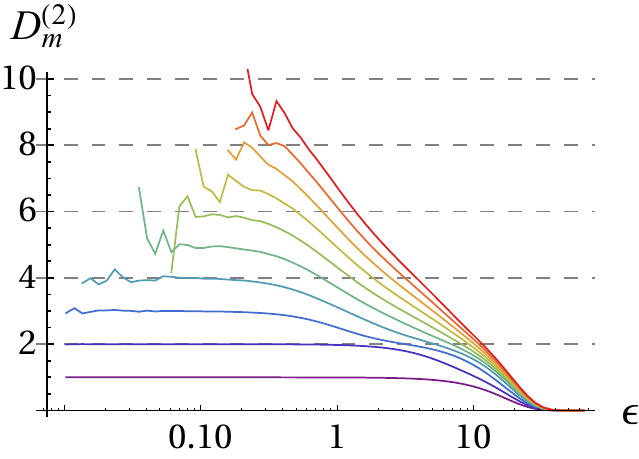}&
    \multirow{2}{*}{\includegraphics[scale=1,trim=0 0 0 -10]{legend10}}\\
    \subplot{a:1:h2} & \subplot{a:10:h2}\\[-1.1em]
    \includegraphics[width=.3\columnwidth]{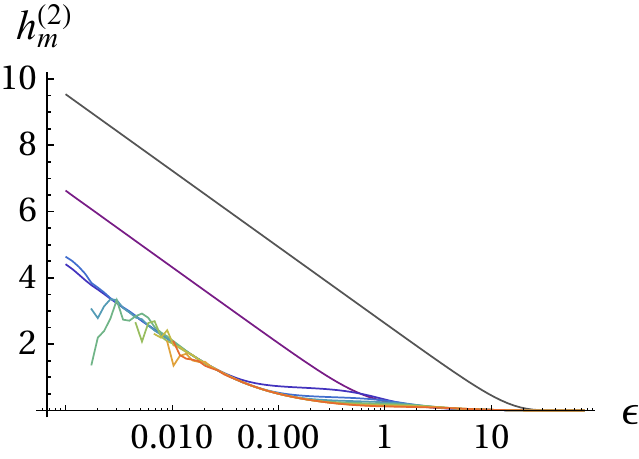}&
    \includegraphics[width=.3\columnwidth]{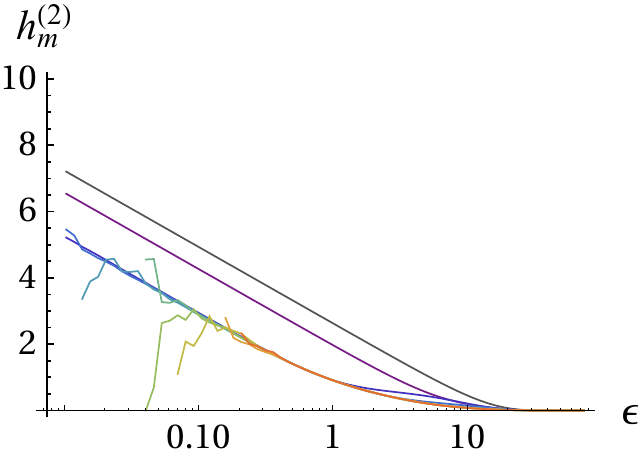}\\
    \subplot{a:1:e2} & \subplot{a:10:e2}\\[-1.1em]
    \includegraphics[width=.3\columnwidth]{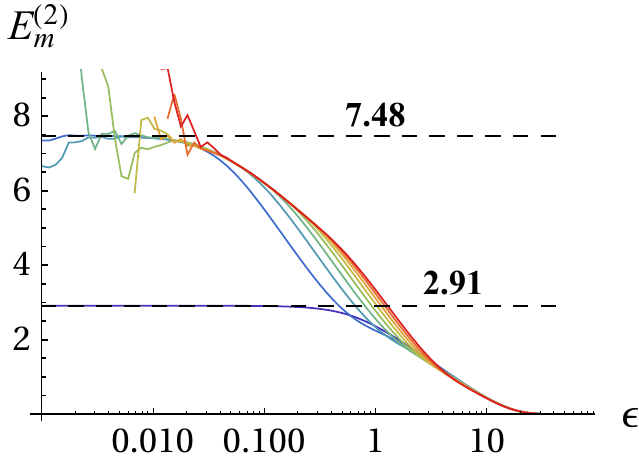}&
    \includegraphics[width=.3\columnwidth]{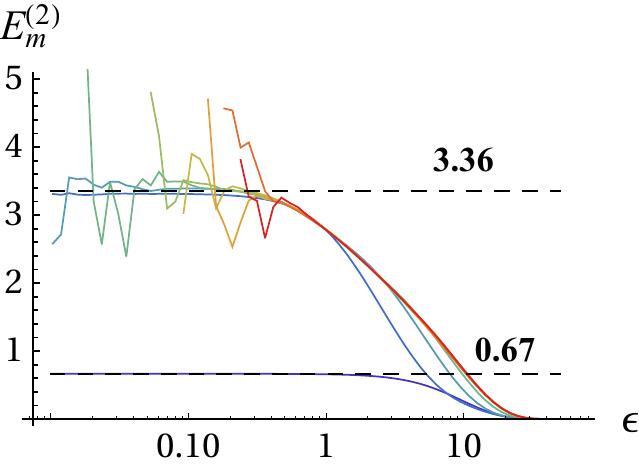}\\
    \subplot{a:1:pi} & \subplot{a:10:pi}\\[-1.1em]
    \includegraphics[width=.3\columnwidth]{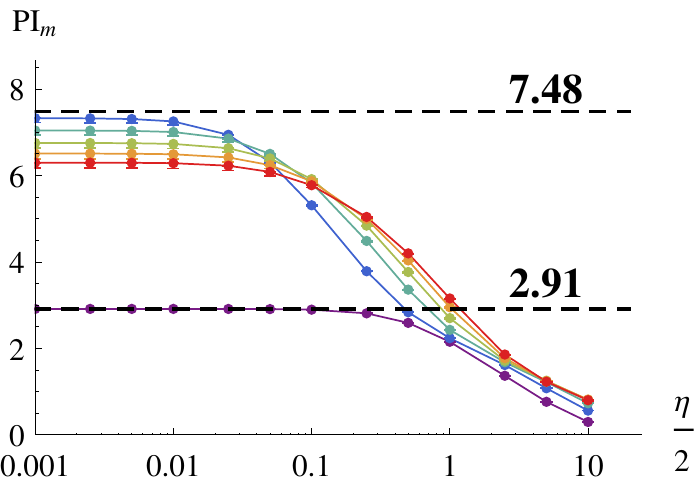}&
    \includegraphics[width=.3\columnwidth]{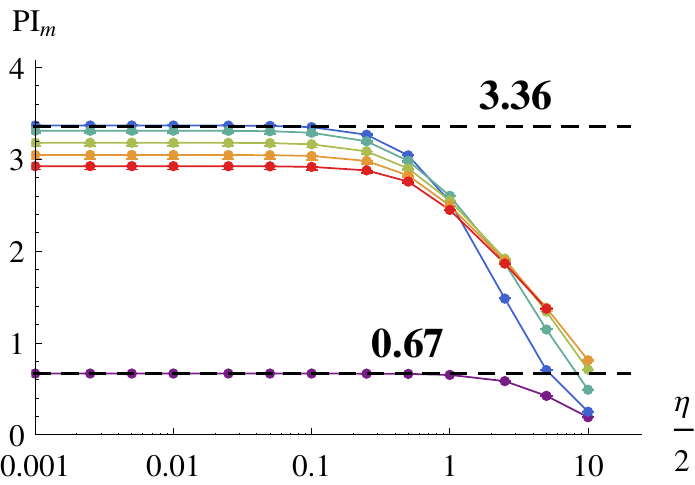}&
    \includegraphics[scale=1,trim=0 -35 0 0]{legend6-1}
  \end{tabular}
  \caption{Analysis of an AR$_2$-model ($a_1 = 1.991843, a_2 = -0.994793$) of the Lorenz attractor observed at delay 1 and delay 10.
    See \fig{fig:lorenz:quant} for details.
    \infig{\ref{a:1:d2}, \ref{a:10:d2}}~The dimensionality corresponds to the embedding dimension (as long as the data suffices).
    \infig{\ref{a:1:h2}, \ref{a:10:h2}}~Conditional entropies show the entropy of the noise for $m>2$. The difference
    between first and second, and second and third embedding reveal the actual structure of the time-series, better visible in \ref{a:1:e2} and \ref{a:10:e2}.
   \infig{\ref{a:1:e2}, \ref{a:10:e2}}~The excess entropy converges to the theoretical values of
   $2.91$ for $m = 1$ and to $7.48$ for $m\ge 2$ (delay 1) and $0.67$ for $m = 1$ and to $3.36$ for $m\ge 2$ (delay 10).
   \infig{\ref{a:1:pi}, \ref{a:10:pi}}~Predictive information estimated with KSG algorithm with $k=20$.
 }   \label{fig:lorenz:armodel}
 \end{figure}

The predictive information \eqnp{eqn:PI} can be also written as
\begin{align}
  \PI_m &= \sum_{k=0}^{m-1} 2 h_k - \sum_{k=0}^{2m-1} h_k\,.
\end{align}
For the AR2 process  we have $h_k=h_2$ for $k\ge 2$ due to Markov order 2. Thus,
$\PI_1 = h_0-h_1=2 H_1 - H_2$ and $\PI_2 = \PI_k = h_0 + h_1 -2h_2 = 3H_2 - 2H_3$ for $k\ge 2$.
For the excess entropy \eqn{eqn:Em} we find $E_1=0$, $E_2=h_0-h_1=\PI_1$, $E_3=E_k=h_0+h_1-2h_2=\PI_2$, for $k\ge 3$.
\Fig{fig:lorenz:armodel} shows the results for the AR2 model obtained from fitting the parameters to the lorenz system ($a_1 = 1.991843, a_2 = -0.994793$).
Indeed the conditional entropies $h_m=h_2$ become identical for $m\ge2$, see \ref{a:1:h2},\ref{a:10:h2}
 and the excess entropies converge for $m\ge 3$ to their theoretical values, see \ref{a:1:e2},\ref{a:10:e2}
In the case of the KSG algorithm the values are correct for $m=1,2$ but for higher embedding dimensions we get a significant underestimation, see \ref{a:1:pi},\ref{a:10:pi}. This effect is even stronger for smaller neighborhood sizes $k$.
\resetsubplot

\section{Details of the Decomposition Algorithm}\label{sec:app:decomp}
\subsection{Fitting}

In deterministic scaling ranges the $\delta h_m(\eps)$ follow the form $o-s \log(\eps)$ (possibly with s=0).
In a perfectly stochastic scaling range we have $\delta h_m(\eps)\propto \textrm{const}$ for all $m$.
However, the behavior of a system may show several deterministic and stochastic scaling ranges.
So we have to determine segments where a certain typical behavior is found.
The following algorithm assumes that for each $m$ the $\delta h_m(\eps)$ contain segments of the form $f(\eps) \approx 0-s \log(\eps)$ and some parts that do not fit.
Because for different $m$ and different systems the curves may have different amount of noise we
first determine a quality threshold for the fits.
Internally the data is represented by $n$ points which we denote $(\eps^i,\delta h^i)$, with $i=1,\ldots,n$.
We denote the range where a fit was computed by $r=(i_l,i_u)$ where $i_l < i_u$.
The quality measure is the sum of the squared fit-residuals:

\begin{align}
  q(r) = \sum_{i\in r}(\delta h(\eps^i) -  f(\eps^i))^2\,.
\end{align}

\begin{figure}
  \centering
  \begin{tabular}{clc}
    (a)&&(b)\\
    \includegraphics[width=.55\columnwidth]{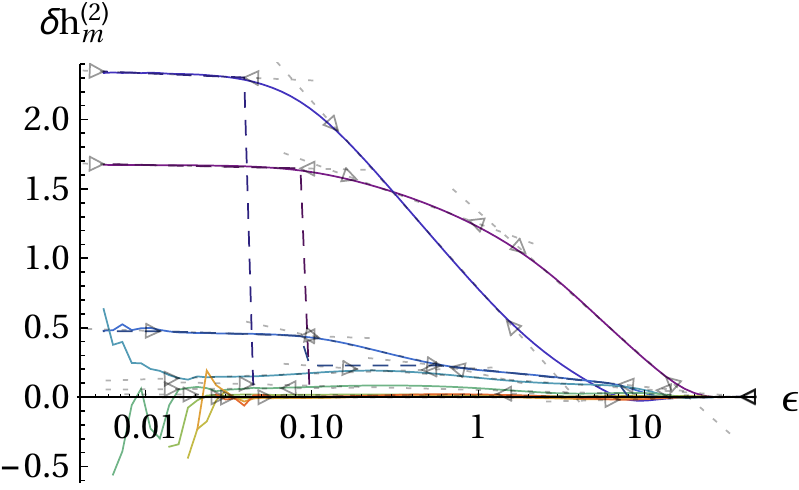}&
    \includegraphics[scale=1.2,trim=0 -35 0 0]{legend10}&
    \includegraphics[width=.3\columnwidth]{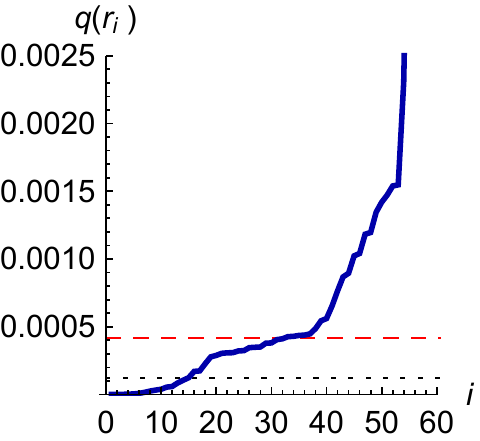}
  \end{tabular}
  \caption{Illustration of the fitting algorithm and the determination of $c_m^{\textrm{MT}}$ and $q_{\textrm{max}}$.
    \infig{(a)}~Shown are the $\delta h_m$ curves for the Lorenz system with 0.005 dynamic noise, see \fig{fig:lorenz:quant}.
    The fits are shown by dotted gray lines and their validity range $r$ is marked by triangles.
    The dashed colored lines show $c_m^{\textrm{MT}}$ for $m=1,2,3$.
    \infig{(b)}~Ranked fit qualities for all segments with 10 data points of $\delta h_1$ (solid line) with $25\%$ percentile (dotted line) and threshold $q_{\textrm{max}}$ (dashed red line). For visibility $y$-axis is cut, $\max q = 0.013$.
  }\label{fig:fitting}
\end{figure}

First, all fits with a certain number of data points (here 10) are computed and their quality measure is computed:
  $Q=\{q(r_j) \mid  |r_j| = 10\}$, where $|r|=i_u-i_l$ is the length of the segment in number of points.
There will be segments with low residual errors and segments with very high residuals, namely those that cover the regions between ideal-typical behavior.
The heuristics for the quality threshold is the $25\%$ percentile of $Q$ plus a small portion of the standard deviation: $q_{\textrm{max}}=P(Q,0.25) + 0.1 \textrm{StdDev(Q)}$.
The reasoning for the percentile is to pick a threshold among the good fits, and we only assume that at least $25\%$ of the curve has fitting segments.
If there is however an extended range of good fits then a high percentage (say $90\%$) of short segments have very similar low $q$ values. In this case a too low threshold would be selected which cuts away good fitting regions.
The tens of the standard deviation is added to lift the threshold above the potential plateau, see \fig{fig:fitting}(b).

For all ``good'' segments $Q_{\textrm{good}} = \{q(r_j) < q_{\textrm{max}} \mid  q(r_j) \in Q\}$ the longest extension of the fitting range is determined that keeps the quality of the fit below the threshold, \ie $\forall q(r_j)\in Q_{\textrm{good}}, (i_l,i_u)=r_j:  i'_u =\argmax_{k>i_u} q((i_l,k))<q_{\textrm{max}}$.
Thus we get many overlapping regions of good fits. For each pair of two regions that overlap more than $30\%$ (w.r.t.~the smaller) we discard the shorter one. The remaining regions are our final fits, as displayed in \fig{fig:fitting}(a).

The components $c^{\textrm{MT}}$ for calculating the constant of the  middle term \eqnp{eqn:E_decomp:algo} are also displayed in \fig{fig:fitting}(a).
 For $m=3$, the value of $c^{\textrm{MT}}$ is $0.23$ for $0.1<\eps<0.8$ because of the shallow region at $[0.8,9]$.
This leads to the value of $0.68$ ($m*c_m^{\textrm{MT}}$) for the state complexity in \fig{fig:lorenz:decomp}\ref{l:n1:de}.
For $\eps<0.1$ the constant $c_3^{\textrm{MT}}$ takes the value of $\approx 0.5$ of the next (smaller scale) plateau.
Similarly $c_{1,2}^{\textrm{MT}}$ raise from zero to the value of the plateaus, which leads to the increase in state-complexity in the stochastic scaling range.
Remember that in the stochastic scaling range ($\eps<0.1$) $m_l=1$ and $m_u=3$ are taken from the preceding deterministic scaling range.

\end{document}